\documentclass[journal]{vgtc}                     

\onlineid{0}

\vgtccategory{Research}

\vgtcpapertype{algorithm/technique}

\title{Residency Octree: A Hybrid Approach for\\Scalable Web-Based Multi-Volume Rendering}

\author{\authororcid{Lukas Herzberger}{0000-0002-9047-065X}, \authororcid{Markus Hadwiger}{0000-0003-1239-4871}, \authororcid{Robert Krüger}{0000-0002-6468-8356}, \authororcid{Peter Sorger}{0000-0002-3364-1838},\\ \authororcid{Hanspeter Pfister}{0000-0002-3620-2582}, \authororcid{Eduard Gr\"oller}{0000-0002-8569-4149}, and \authororcid{Johanna Beyer}{0000-0002-3505-9171}}

\authorfooter{
\item
 Lukas Herzberger and Eduard Gr\"oller are with TU Wien. E-mail: \{herzberger, groeller\}@cg.tuwien.ac.at.
\item
 Markus Hadwiger is with King Abdullah University of Science and Technology (KAUST). E-mail: markus.hadwiger@kaust.edu.sa.
\item
 Johanna Beyer, Robert Krueger, and Hanspeter Pfister are with the John A. Paulson School of Engineering and Applied Sciences at Harvard University. E-mail: \{jbeyer, krueger\}@g.harvard.edu, pfister@seas.harvard.edu.
\item
 Peter Sorger and Robert Krueger are with Harvard Medical School. E-mail: {peter\_sorger, robert\_krueger}@hms.harvard.edu.
}

\abstract{We present a hybrid multi-volume rendering approach based on a novel \textit{Residency Octree} that combines the advantages of out-of-core volume rendering using page tables with those of standard octrees. 
Octree approaches work by performing hierarchical tree traversal.
However, in octree volume rendering, tree traversal and the selection of data resolution are intrinsically coupled. This makes fine-grained empty-space skipping costly. 
Page tables, on the other hand, allow access to any cached brick from any resolution. However, they do not offer a clear and efficient strategy for substituting missing high-resolution data with lower-resolution data.
We enable flexible mixed-resolution out-of-core multi-volume rendering by decoupling the cache residency of multi-resolution data from a resolution-independent spatial subdivision determined by the tree.
Instead of one-to-one node-to-brick correspondences, each residency octree node is mapped to a set of bricks from different resolution levels.
This makes it possible to efficiently and adaptively choose and mix resolutions, adapt sampling rates, and compensate for cache misses.
At the same time, residency octrees
support fine-grained empty-space skipping, independent of the data subdivision used for caching.
Finally, to facilitate collaboration and outreach, and to eliminate local data storage, our implementation is a web-based, pure client-side renderer using WebGPU and WebAssembly.
Our method is faster than prior approaches and efficient for many data channels with a flexible and adaptive choice of data resolution.
}


\teaser{
  \centering
  \includegraphics[width=\linewidth]{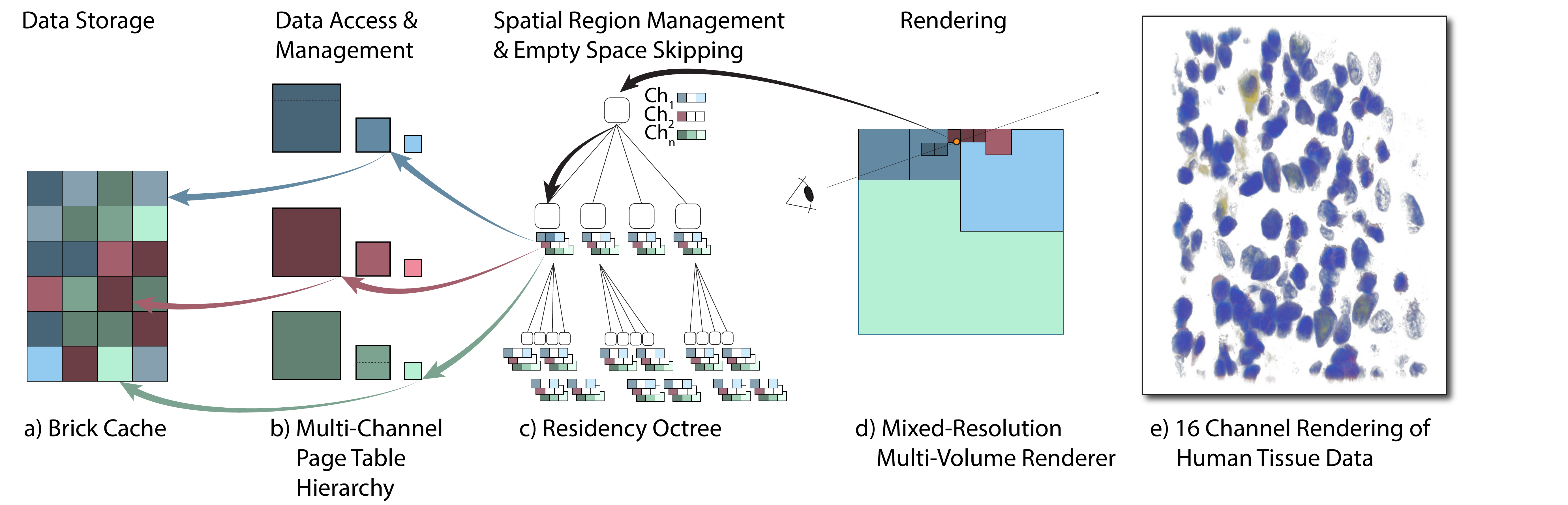}
  \vspace{-7mm}
  \caption{\textbf{Overview.} Volume bricks of different channels (red, green, blue) and resolution levels (dark to bright) are streamed into a brick cache (a), referenced by a multi-channel page table hierarchy (b). Our \emph{residency octree} (c) keeps track of the correspondence between spatial regions and the cache residency of bricks of different resolutions, enabling mixed-resolution, multi-channel rendering (d) with efficient, adaptive substitution of missing higher resolutions by available lower resolutions. Ray traversal is done for spatial regions corresponding to octree nodes instead of bricks, and is independent of the number of channels. (e) 16-channel rendering of melanoma.}
  \label{fig:teaser}
}

\graphicspath{{figs/}{figures/}{pictures/}{images/}{./}} 

\usepackage{tabu}                      
\usepackage{booktabs}                  
\usepackage{lipsum}                    
\usepackage{mwe}                       

\usepackage{mathptmx}                  

\usepackage[ruled,linesnumbered]{algorithm2e}
\usepackage{tabularray}
\usepackage{makecell}
\usepackage{comment}
\usepackage{caption}
\usepackage{subcaption}

\begin{document}


\firstsection{Introduction}
\label{sec:introduction}

\maketitle

Recent advances in imaging modalities produce large-scale volumetric data sets with a large number of channels that require out-of-core methods for direct volume rendering (DVR), such as octrees or page table hierarchies.
An example of such large-scale data sets is Immunofluorescence (IF) imaging data. IF technologies, such as CyCIF~\cite{lin_highly_2018}, are used in the field of digital histopathology to image biological tissue.
The resulting multiplexed images contain information for millions of cells in up to 60 channels, where each channel represents the response of cells to one or more marker antibodies~\cite{krueger_facetto_2020,rashid_narrative_2022}, making proteins visible and thereby revealing the types and functions of cells.

File sizes of IF data sets range from gigabytes to terabytes, and continue to grow as sample sizes, the number of recorded channels, and imaging resolutions increase~\cite{lin_multiplexed_2023}.
File sizes often exceed the available memory. Therefore, out-of-core techniques like bricking, multi-resolution hierarchies, and ray-guided rendering approaches, as discussed by Beyer et al.~\cite{beyer_state---art_2015}, are necessary to visualize such data.

Since it is common for multiple people to collaborate on such data using a heterogeneous set of tools and applications~\cite{krueger_facetto_2020,jessup_scope2screen_2022,rashid_narrative_2022}, it is desirable to provide the data via a web server and use web-based visualization tools instead of relying on native applications.
Due to limitations such as the lack of general-purpose computing on the GPU (GPGPU) in WebGL 2.0, previous web-based volume rendering research has often focused on minimizing the effects of network latency~\cite{yang_volumetric_2015,mwalongo_web-based_2018,adochiei_web_2019} as well as on optimizing the rendering performance itself by offloading
some or all
rendering work to a dedicated server~\cite{wangkaoom_high-quality_2015,qiao_html5-based_2017,raji_scalable_2017}.
However, with the emergence of WebAssembly~\cite{rossberg_webassembly_2019} and WebGPU~\cite{maxfield_myles_webgpu_2023}, web-based scientific visualization applications are now comparable to native applications both in terms of performance and development effort, as, e.g., shown by Usher and Pascucci~\cite{usher_interactive_2020}. These developments now make it feasible to design out-of-core volume rendering algorithms for the web that are similar to those developed for native applications.

Apart from the memory pressure introduced by large file sizes, another problem that arises when visualizing highly multiplexed data sets is the large number of different channels.
Even though, in practice, only a subset of $m \le n$ channels (e.g., $m = 4$) out of all $n$ available channels is visualized at a time, rendering more than one channel using DVR requires careful optimization due to the performance impact of accessing each sample position in multiple volumes.
A common optimization in DVR is empty-space skipping, where empty regions, e.g., those that are fully transparent according to the current transfer function, in the volume are not sampled in order to reduce the number of loop iterations and texture look-ups during rendering~\cite{zellmann_rapid_2018,zellmann_linear_2019,zellmann_hybrid_2019,zellmann_binned_2021,fernandes_bucket_2020,liu_octree_2013,hadwiger_sparseleap_2018,deakin_accelerated_2019,deakin_efficient_2020,xue_parallel_2019,brix_visualization_2014}.
However, most existing web-based volume renderers are neither designed for large-scale multi-channel data nor do they use optimizations such as fine-grained empty-space skipping techniques~\cite{manz_viv_2022,google_neuroglancer_2022}.

The subset of visualized channels is usually user-defined and may change at run-time.
For this reason, acceleration structures used to optimize rendering performance need to be flexible enough to allow for channel selection switches.
Furthermore, channels are not necessarily equally important, for example, by having different frequency content, or they are simply less interesting to the user in the current context.
This makes it desirable to render more important channels in higher resolution while rendering less important channels in lower resolution in order to reduce the memory required for storing the currently visible volume data on the GPU.
Existing techniques for multi-volume data for native environments are not designed for channel switches and do not support rendering different channels at different resolutions~\cite{brix_visualization_2014,drees_voreen_2022}.

Ray-guided DVR methods have been shown to work best for large-scale volumetric data~\cite{beyer_state---art_2015}.
They can be divided into two families: page-table-based and octree-based approaches.
The former allows direct access to any cached brick from any resolution level.
In octree-based approaches, each node represents exactly one brick in the data set.
Because of the hierarchical tree structure, such approaches enforce the traversal algorithm used to access cached volume data during rendering as well as the order in which bricks of different resolutions are streamed in, i.e., from the root node to leaf nodes.
This makes page-table-based approaches more flexible than octree-based ones, but they do not offer a clear and efficient strategy for substituting missing high-resolution data with lower-resolution data guaranteed to be resident in the cache~\cite{beyer_state---art_2015}.
Both families allow for empty-space skipping by either skipping over empty bricks or empty octree nodes, respectively.

\begin{figure}[tb]
 \centering
  \includegraphics[width=1\linewidth]{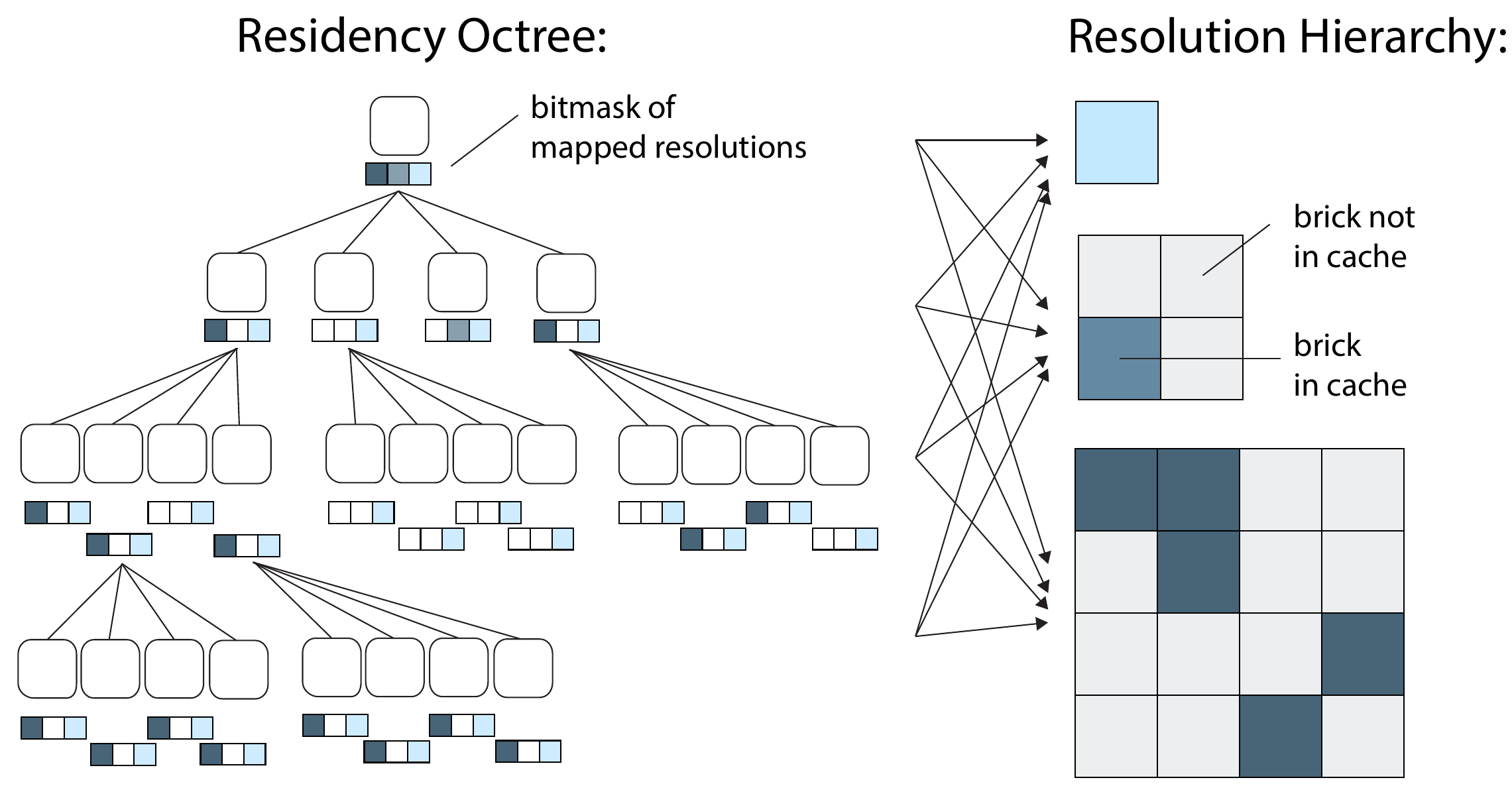}
 \caption{\textbf{Residency octree nodes} store cache residency information for sets of corresponding bricks of each resolution level in bitmasks. Here, we focus on a single-channel volume; brick colors indicate the resolution level. For fine-grained empty-space skipping, the number of levels of the residency octree (here, 4) can be independent of the resolution hierarchy.}
\vspace{-4mm}
   \label{fig:residency_octree}
 \end{figure}
 
\begin{figure*}[tb]
 \centering
  \includegraphics[width=1\linewidth]{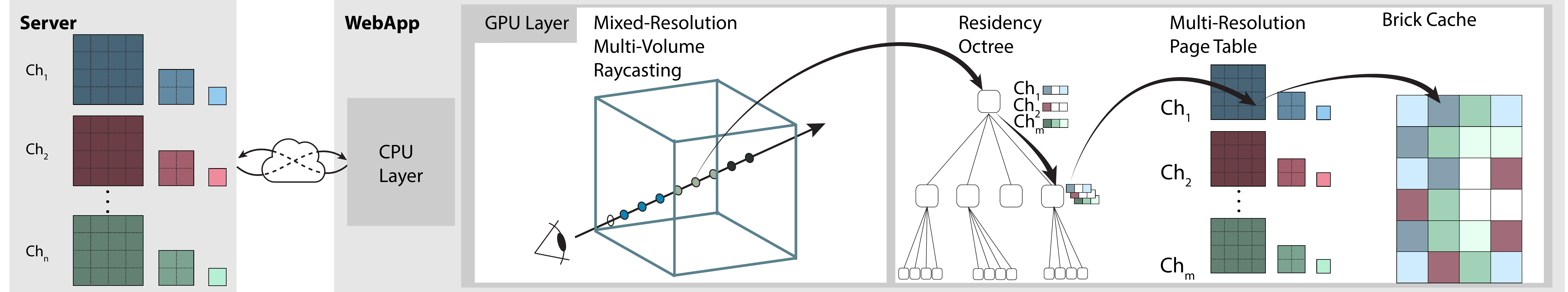}
 \caption{\textbf{System overview.} A bricked multi-resolution multi-volume hierarchy with $n$ channels is provided by one or more file servers. Client-side mixed-resolution multi-volume rendering traverses the volume via the residency octree. Each octree node contains $m \leq n$ bitmasks corresponding to the in-cache availability of different resolutions for $m$ channels. Actual data access uses a multi-resolution multi-channel page table, with bricks streamed into a cache. The residency octree enables efficiently mixing channel resolutions and substituting alternative resolutions on cache misses.}
\vspace{-4mm}
   \label{fig:system_overview}
 \end{figure*}
 
\textbf{Residency Octree.} We propose a novel residency octree, a hybrid data structure for out-of-core DVR of multi-volume data that combines the advantages of page tables with those of octrees.
It decouples the cache residency of multi-resolution data from a resolution-independent spatial subdivision determined by the tree.
\cref{fig:residency_octree} depicts a residency octree for a single-channel volume.
Instead of one-to-one node-to-brick correspondences as in standard octrees, each residency octree node represents a resolution-independent spatial region in the volume that references all resolution levels in a bricked volume hierarchy.

This makes it possible to efficiently and adaptively choose and mix resolutions, adapt sampling rates, and compensate for cache misses.
At the same time, this decoupling allows residency octrees to support fine-grained empty-space skipping, independent of the data subdivision used for bricking.
For this purpose, each residency octree node stores transfer-function independent metadata, e.g., minimum and maximum scalar values in the spatial region represented by the node, alongside the information about resolution levels in which the node's corresponding bricks are currently resident in the cache.
Internally, our data structure is backed by a multi-channel page table hierarchy and brick cache.
By keeping information about multiple resolutions and channels in each node, our data structure works well for multi-channel empty-space skipping and allows mixing different resolutions not only for single-channel but also for multi-channel data.
Our data structure has been fully implemented on the GPU and facilitates efficient run-time changes to the selection of visible channels.
With WebGPU~\cite{maxfield_myles_webgpu_2023}, our method enables pure client-side out-of-core multi-volume rendering on the web.
\cref{fig:teaser} (left) and \cref{fig:system_overview} give an overview of our system architecture.

\textbf{Contributions.}
(1) A novel hybrid data structure for out-of-core volume rendering of multi-volume data sets that decouples the resolution levels in a bricked volume hierarchy from the spatial subdivision determined by the tree.
This decoupling makes it possible to efficiently and adaptively choose and mix resolutions, both between samples of a single channel and samples taken from multiple channels, adapt sampling rates accordingly, compensate for cache misses, and skip empty space.
(2) A mixed-resolution multi-volume rendering algorithm to visualize multiple channels at different resolutions and dynamically take into account differences in channel importance.

\section{Related Work}
\label{sec:related-work}

\textbf{Large-scale volume rendering.}
\begin{figure}[b!]
\vspace{-4mm}
 \centering
 \includegraphics[width=0.9\columnwidth]{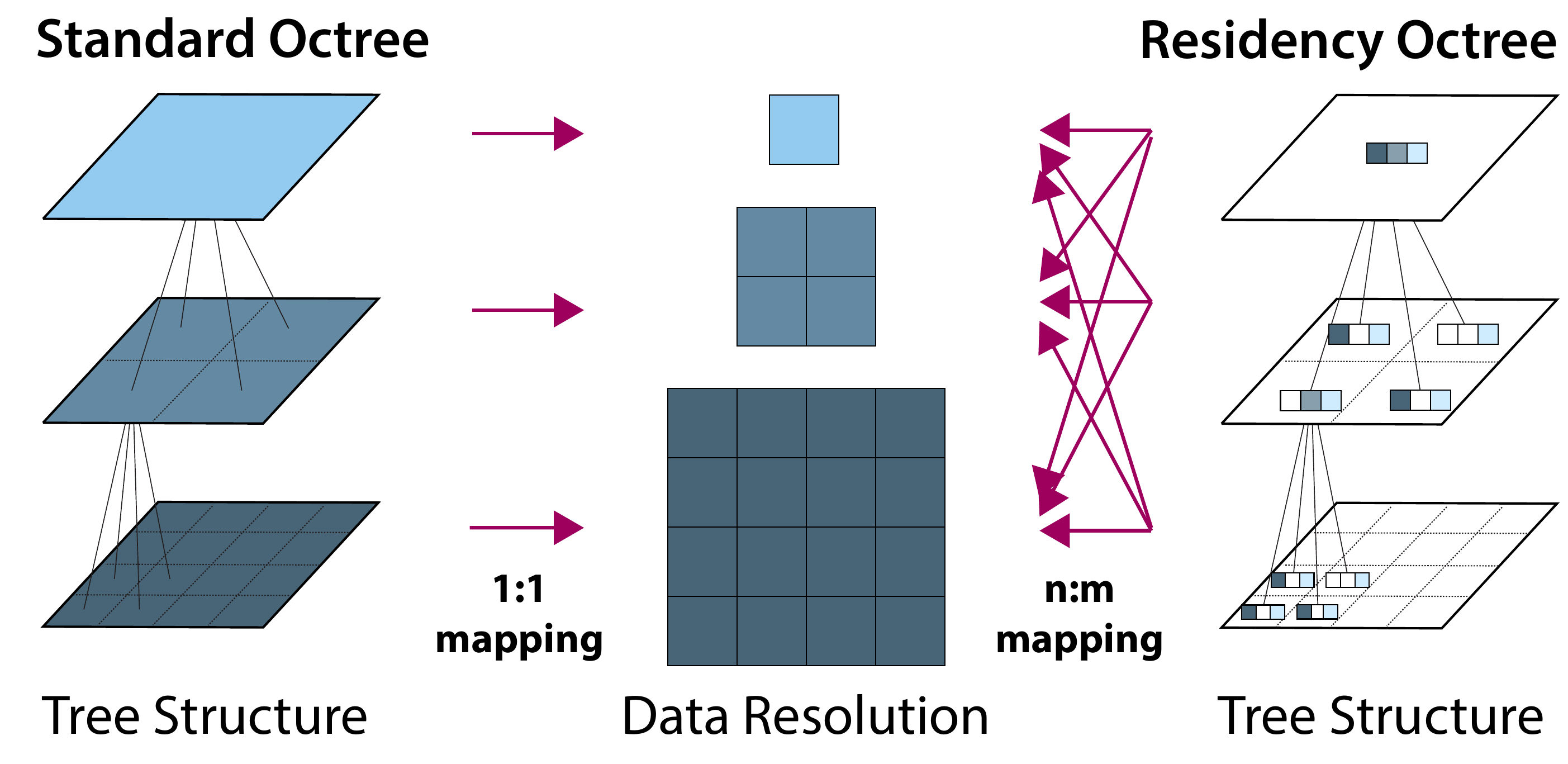}
 \caption{\textbf{Previous octree-based out-of-core approaches} (left) employ a one-to-one mapping between octree nodes and bricks. In contrast, the residency octree nodes in our approach (right) represent geometric spatial regions, with each node mapping to multiple bricks and vice versa.}
 \label{fig:previous-vs-ours}
\end{figure}
Childs defines data as large if they are ``too large to be processed ... (1) in its entirety, (2) all at once, and (3) exceeds the available memory''~\cite[p.~9]{bethel_high_2012}.
Following this definition, Beyer et al.~\cite{beyer_state---art_2015} define scalability of GPU-based large-scale visualization approaches in terms of the minimal subset of the data required to render an image at a desired resolution - the \textit{working set}.
A visualization approach is considered scalable if the working set size depends only on the output resolution and the data visible on the screen but is independent of the size of the whole data set.
Several scalable volume rendering techniques have been developed for desktop environments~\cite{beyer_state---art_2015}.
For these approaches, the data set is (a) present in a multi-resolution hierarchy and (b) split into smaller bricks where all bricks across all resolutions have the same size in voxels.
In ray-guided volume rendering~\cite{crassin_gigavoxels_2009,hadwiger_interactive_2012,fogal_analysis_2013,brix_visualization_2014,sarton_interactive_2020}, the working set is determined during the ray casting process itself by recording which bricks were traversed by viewing rays.
In each frame, a list of brick requests is compiled to stream in missing volume data on demand.
In octree-based ray-guided rendering, the downsampling ratio between resolution levels in the multi-resolution hierarchy is intrinsically coupled to the spatial subdivision determined by the tree structure such that each node corresponds to exactly one brick in the data set~\cite{crassin_gigavoxels_2009,brix_visualization_2014}.
Crassin et al.~\cite{crassin_gigavoxels_2009} use a node pool to keep subtrees resident in the cache.
In turn, each node in the node pool references volume data stored in a brick cache.
Brix et al.~\cite{brix_visualization_2014} extend this structure to multi-channel data.

Hadwiger et al.~\cite{hadwiger_interactive_2012} propose a memory virtualization technique based on page tables instead of a tree structure.
Their method supports arbitrary downsampling ratios for the data set and accessing volume data of the desired resolution directly.
However, their technique does not give a clear strategy for substituting missing bricks by other resolutions resident in the cache.
Fogal et al.~\cite{fogal_analysis_2013} propose a similar approach but use a global hash table for reporting cache misses back to the CPU instead of one per image tile.
Sarton et al.~\cite{sarton_interactive_2020} generalize the concept of page tables for volumetric data sets for non-rendering purposes.
They use CUDA shared memory to process cache misses on the GPU itself to minimize memory transfers between the CPU and the GPU.

Our approach combines the advantages of both octrees and page table hierarchies by decoupling the cache residency of multi-resolution data from the spatial subdivision determined by the residency octree.
\cref{fig:previous-vs-ours} illustrates how our novel approach differs from previous octree-based methods.
Instead of having one-to-one correspondences between bricks and nodes, our approach decouples resolution levels in the data set and the spatial subdivisions determined by our octree structure such that each node maps to one or more bricks in each resolution level.

\textbf{Multi-volume rendering}
considers multiple co-registered volumes at once and within one view, evaluating each sample along a viewing ray for all currently visible channels.
Schubert and Scholl~\cite{schubert_comparing_2011} call this \textit{accumulation level intermixing}.
Brix et al.~\cite{brix_visualization_2014} present an out-of-core rendering approach for multi-channel volume data sets that is based on Crassin et al.'s~\cite{crassin_gigavoxels_2009} octree-based memory virtualization scheme. Their approach is implemented in Voreen~\cite{drees_voreen_2022}.
Viv~\cite{manz_viv_2022} is a web-based renderer that stores each channel in a separate texture.
The number of channels that can be represented simultaneously on the GPU is therefore limited only by the number of available texture bind points allowed by the API, e.g., WebGL 2.0 guarantees a minimum of eight texture bind points.
Channels may be switched at run-time by simply changing the texture bindings in the shader.
Neuroglancer~\cite{google_neuroglancer_2022} uses the same technique for representing multiple channels on the GPU.

\textbf{Empty-space skipping.}
One of the major problems with DVR is its high computational cost due to evaluating hundreds to thousands of samples per ray in each frame.
Empty-space skipping methods accelerate this process by determining regions of empty space that do not require extensive sampling, effectively reducing the number of samples that have to be evaluated.
For this purpose, volumes are subdivided into smaller parts, often involving a tree-like hierarchy, such as octrees~\cite{crassin_gigavoxels_2009,liu_octree_2013,brix_visualization_2014,faludi_transfer-function-independent_2022}, kd-trees~\cite{zellmann_rapid_2018,zellmann_hybrid_2019,zellmann_binned_2021}, or linear bounding volume hierarchies (LBVH)~\cite{zellmann_linear_2019,fernandes_bucket_2020}.
In recent years, several parallel tree construction algorithms have been proposed to support run-time re-construction of acceleration data structures, e.g., as a result of changes to the transfer function used~\cite{zellmann_rapid_2018,zellmann_linear_2019,fernandes_bucket_2020,zellmann_binned_2021}.
Other approaches include the use of Chebychev distance maps~\cite{deakin_accelerated_2019,deakin_efficient_2020} or subdividing the volume into non-uniform grids~\cite{xue_parallel_2019}.
Our approach builds on an octree structure that is incrementally constructed on the GPU.

Instead of traversing a tree structure on the GPU, Liu et al.~\cite{liu_octree_2013} propose rasterizing proxy geometry representing a view-dependent cut of octree nodes.
Hadwiger et al.~\cite{hadwiger_sparseleap_2018} also utilize rasterization hardware to avoid hierarchy traversal on the GPU by only traversing per-pixel ray segment lists produced through rasterizing occupancy geometry.
Wang et al.~\cite{wang_composition-free_2021} use a tile-based rendering scheme combined with an octree structure with a fixed leaf node size to render per-tile node lists.

Many acceleration structures store transfer function dependent visibility information and thus require reconstruction after transfer function changes~\cite{zellmann_rapid_2018,zellmann_linear_2019,zellmann_hybrid_2019,fernandes_bucket_2020,deakin_accelerated_2019,deakin_efficient_2020,zellmann_binned_2021}. This is not the case when transfer function independent information is stored, such as minimum and maximum values or a histogram of values contained in spatial regions in the volume as proposed by Faludi et al.~\cite{faludi_transfer-function-independent_2022}.
Residency octree nodes also store transfer function independent culling data for empty-space skipping.

\textbf{Web-based volume rendering.}
In web-based environments, volume rendering research is mostly focused on solutions to network latency when loading volumetric data~\cite{yang_volumetric_2015,yang_volumetric_2015,mwalongo_web-based_2018,adochiei_web_2019} and improving volume rendering performance on the web, e.g., by offloading
rendering work to remote rendering servers
~\cite{fogal_large_2010,beyer_distributed_2011,wangkaoom_high-quality_2015,qiao_html5-based_2017,raji_scalable_2017,sarton_distributed_2019}.

In contrast to server-side rendering, pure client-side renderers only require a file server as a backend 
\cite{congote_interactive_2011,mobeen_high-performance_2012,diaz-garcia_progressive_2018,arbelaiz_progressive_2019,manz_viv_2022,google_neuroglancer_2022}.
To reduce the data size streamed from servers to client-side applications, Moore et al.~\cite{moore_ome-ngff_2021} present OME-NGFF, an open file format for bricked data sets. It supports fast access to individual bricks of the data set by storing them in separate compressed files. This also makes it useful for use in a cloud-based context since compressed bricks may be distributed across multiple file servers.
Manz et al.~\cite{manz_viv_2022} present Viv, a WebGL-based library for visualizing biomedical data.
While their library focuses on 2D data, it also allows DVR of 3D data.
However, it is limited to the available GPU memory and thus does not support large-scale data.
Neuroglancer~\cite{google_neuroglancer_2022} is a WebGL-based tool for visualizing and annotating large-scale volumetric data.
It uses a bricked multi-resolution hierarchy approach for its DVR view to scale to large data sets.
In the DVR view, all bricks are selected from the same resolution based on global camera parameters, leading to suboptimal working sets.
Current web-based volume rendering solutions are limited by WebGL's lack of compute pipelines and general shader storage buffers.
Our implementation (\cref{sec:implementation}) uses the upcoming WebGPU standard, which provides a range of algorithms proposed for native environments to be implemented in web contexts~\cite{usher_interactive_2020}, and OME-Zarr, an implementation of OME-NGFF, for representing bricked multi-resolution volume data, with LZ4-compressed chunks for efficient data transfer over a network.

\textbf{Texture compression.} 
An important aspect of out-of-core volume rendering is 3D texture compression, which reduces the memory footprint of the volume texture data.
Examples are compression domain volume rendering~\cite{schneider_compression_2003}, e.g., during ray casting~\cite{kruger_acceleration_2003}, using hardware texture compression methods such as ASTC~\cite{nystad_adaptive_2012}, or other fixed-rate compressors~\cite{lindstrom_fixed-rate_2014}, to enable massive volume visualization~\cite{marton_framework_2019}. We refer to the state-of-the-art report by Balsa Rodríguez et al.~\cite{balsa_rodriguez_state---art_2014} for a more extensive review of the literature.
Since WebGPU currently does not support compressed 3D textures natively~\cite{maxfield_myles_webgpu_2023}, our current renderer does not use compressed textures. However, we use LZ4-compressed bricks for efficient data transfer over the network.
We consider hardware texture compression an orthogonal approach to our octree data structure and leave its integration to future work.

\section{Basics and Terminology}
\label{sec:terminology}

We make a clear distinction between the resolution levels of a volume and the spatial, purely geometric subdivision determined by the residency octree, i.e., the geometric boundaries of octree nodes.

\textbf{Brick hierarchy.}
Each data set is represented as a bricked volume hierarchy consisting of multiple \emph{resolution levels}.
A \emph{brick} in this hierarchy is a chunk of voxel data belonging to a specific resolution level.
Each brick has a \emph{size in voxels} (e.g., $32^3$) that is the same for all resolution levels, and that is therefore independent of its \emph{spatial extent}. The latter is determined by the resolution level the brick belongs to.
To access a brick's data on the GPU, it has to be \emph{resident} in a \emph{brick cache}.

\begin{figure}[tb]
 \centering
 \includegraphics[width=\columnwidth]{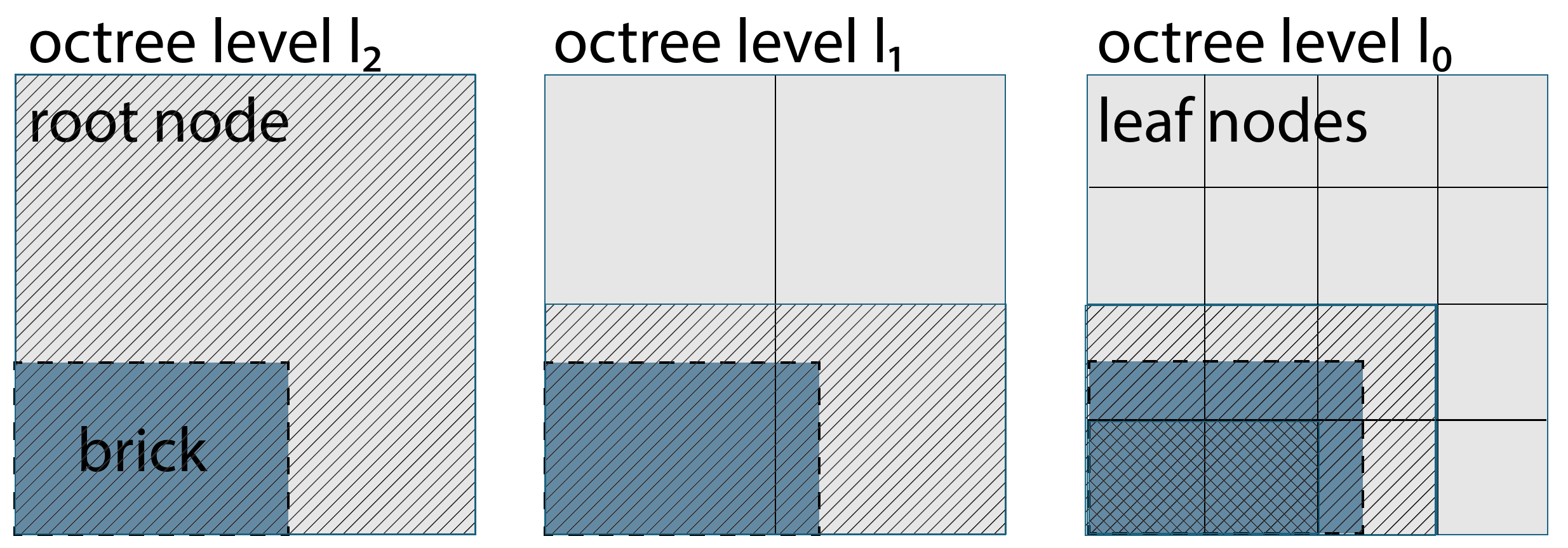}
 \caption{\textbf{Bricks decoupled from residency octree nodes.} A brick (blue; boundary indicated by dashed lines) maps to nodes in all octree levels (octree node boundaries indicated by solid lines). Some nodes are only \emph{partially mapped} (hatched) when this brick is resident in the cache, while others are \emph{fully mapped} (cross-hatched) in the brick's resolution level.}
\vspace{-4mm}
  \label{fig:brick-vs-node}
\end{figure}

\textbf{Residency octree hierarchy.}
In contrast, a residency octree has multiple \emph{subdivision levels}.
While the downsampling ratio between resolution levels in the bricked volume hierarchy is flexible (see, e.g., \cite{hadwiger_interactive_2012}), the spatial subdivision determined by the octree is not.
A \emph{node} in the residency octree represents a region in the volume whose spatial extent is determined by the subdivision level it belongs to.
Other than in previous work~\cite{crassin_gigavoxels_2009,brix_visualization_2014}, we do not associate a size in voxels with a single octree node, as illustrated in \cref{fig:previous-vs-ours}.
In previous octree-based out-of-core DVR approaches, the resolution levels of the bricked volume hierarchy are intrinsically coupled to the subdivision levels of the octree due to a one-to-one mapping between bricks and octree nodes.

In our approach, these two concepts are fully decoupled, such that an octree node corresponds to a \emph{set} of bricks in each resolution level in the bricked volume hierarchy.
The number of bricks of a resolution level required to fully cover the spatial extent of an octree node depends on the spatial extent of bricks in that resolution level, which differs from the spatial extent of the octree node.
For instance, the spatial extent of a single brick in the lowest resolution level may cover the entire volume space and thus cover all residency octree nodes.
Vice versa, a brick in a higher resolution level will only cover a part of the residency octree's root node because the latter covers the whole volume.

\textbf{Fully vs.\ partially mapped residency octree nodes.}
When for a given octree node, the cache-resident bricks of a given resolution level fully cover the node's spatial extent, we refer to the node as \emph{fully mapped} in that resolution level.
Similarly, if at least one brick of a resolution level whose spatial extent overlaps with a node's spatial extent is resident in the cache, the node is \emph{partially mapped} in that resolution level.
This implies that a node that is fully mapped in a resolution level is also partially mapped, but not vice versa.
Furthermore, if a node is fully mapped in some resolution level, all of its child nodes are also fully mapped in that resolution level; and if a node is partially mapped in some resolution level, its parent node is too.
Naturally, as \cref{fig:brick-vs-node} illustrates, the same brick can fully map some nodes, while only partially mapping others. A residency octree node may be partially and fully mapped in different resolutions, respectively, at the same time.

\textbf{Multi-channel data.}
If a volume has multiple channels, the bricked volume hierarchy is extended to a \emph{bricked multi-volume hierarchy}, where we assume all channels to use the same brick size in voxels.
The relationship of octree nodes and bricks in this multi-volume hierarchy is similar to the single-channel case, but instead of having a single set of corresponding bricks per resolution level, a node now has one such set for each channel and resolution level.

\section{System Overview}
\label{sec:system-overview}

Our system is illustrated in \cref{fig:teaser} (left) and \cref{fig:system_overview}.
The bricks in the bricked multi-volume hierarchy are provided by a server.
Depending on the implementation, the server may offer file storage, or it may have some additional capabilities, e.g., querying metadata for specific regions in the volume. Examples are minimum and maximum scalar values or a histogram of values within a region in the volume.

Bricks and metadata are consumed by a client-side multi-volume rendering application that uses a ray-guided renderer to determine which data to fetch from the server.
It consists of two main parts: the novel Residency Octree presented in detail in \cref{sec:residency-octree}, and the mixed-resolution multi-volume renderer discussed in \cref{sec:mixed-resolution-multi-volume-renderer}.

The residency octree manages information about volume data on the GPU by keeping track of which bricks that are currently resident in the brick cache map to which octree nodes.
It does this in conjunction with a page table hierarchy similar to those presented in previous work~\cite{hadwiger_interactive_2012,sarton_interactive_2020}, extended to multiple channels as discussed in \cref{sec:memory-management}.
Additionally, the residency octree provides transfer-function independent metadata for each node, which the renderer uses for empty-space skipping.
For multi-volume data with $n$ different channels, the residency octree can represent $m \leq n$ channels at a time.
The maximum number $m$ of visible channels is user-defined and set at initialization time.
However, the mapping of channels in the data set to channels referenced by our hybrid data structure can change at run-time.

The mixed-resolution multi-volume renderer accesses both metadata and cached volume data through the residency octree.
If either is missing for a node, it generates a request for the missing data to be streamed in from the server.
Since these requests are generated by ray traversal during rendering our method is ray-guided.

Both the residency octree and the mixed-resolution multi-volume renderer are fully implemented on the GPU.
The CPU is mainly needed for driving the overall rendering. It dispatches rendering and memory management commands to the GPU, forwards brick requests to the server, and uploads bricks received from the server to the GPU.
In order to keep buffer copies from the GPU to the CPU small, we specify a global limit on the number of recorded cache misses per frame.

\section{Residency Octree and Out-of-Core Rendering}
\label{sec:residency-octree}

\cref{fig:teaser}~(c) illustrates a residency octree for a multi-channel volume, and \cref{fig:residency_octree} highlights aspects using a single-channel volume.
Bricks of different channels and resolution levels are all stored in the same brick cache, which is referenced by a multi-channel page table hierarchy discussed in \cref{sec:memory-management}.
The residency octree stores both transfer-function independent metadata and brick cache residency information, keeping track of the resolution levels each node is currently mapped in.
This makes it possible to efficiently substitute missing bricks in the desired resolution with bricks of another resolution without requiring the bricks of all lower resolutions to be resident in the cache.
In \cref{sec:octree}, we discuss the layout of residency octree nodes, and in \cref{sec:octree-updates}, we discuss how residency octree nodes are updated when new data is streamed in.

\subsection{Multi-channel multi-resolution page table hierarchy}
\label{sec:memory-management}

We use a multi-resolution page table hierarchy to manage volume data as in earlier work~\cite{hadwiger_interactive_2012}, fully implemented in GPU kernels~\cite{sarton_interactive_2020}, extended to multiple channels, and implemented in WebGPU.
A single brick cache stores the working set on the GPU, referenced by a multi-resolution page table hierarchy for each channel.
This virtualizes a bricked multi-resolution volume via a paged 3D address space with page tables storing all information required for virtual-to-physical address translation, and keeping track of the cache residency of bricks.
Each page table hierarchy represents one resolution level in the volume hierarchy~\cite{hadwiger_interactive_2012}, referencing all bricks comprising that resolution level.

Like previous out-of-core renderers~\cite{hadwiger_interactive_2012,fogal_analysis_2013,sarton_interactive_2020}, all bricks in the multi-volume hierarchy have the same size in voxels, even though they can have different spatial extents in the volume, depending on the resolution level they belong to.
This greatly simplifies cache management since each entry in the brick cache has the same size.
For multi-channel volumes, although we employ one multi-resolution page table hierarchy per channel, all channels still share the same brick size and cache.

To address voxel data, we use virtual multi-resolution addresses $(l,\textbf{p})$~\cite{hadwiger_interactive_2012}, where $l$ is the resolution level and $\textbf{p} \in [0,1]^3$ are the normalized floating-point coordinates (e.g., a ray sample's coordinates) in the page table hierarchy's reference space.
In order to address a position in a multi-volume hierarchy, we extend this to a virtual multi-resolution, multi-channel address $(l,c,\textbf{p})$, where $c$ is the integer channel index.

To efficiently communicate cache misses recorded during rendering to the CPU, a unique integer ID is assigned to each brick, encoding the position and resolution level in the bricked volume~\cite{hadwiger_interactive_2012,fogal_analysis_2013,sarton_interactive_2020}.
To address multi-channel volumes, we additionally encode each brick's channel index in its ID.
As shown in earlier work~\cite{hadwiger_interactive_2012}, the bit-width of the data type used for brick IDs, e.g., 32-bit or 64-bit integers, constrains the maximum size of a volume.
Since we also encode the channel index into an ID, the maximum number of channels that can be represented by a multi-channel page table hierarchy is also constrained by the bit-width of the data type used.
In order to still support more channels than can be represented by the multi-channel page table hierarchy, we store a mapping of the channels currently used by our data structure to the channels in the multi-volume hierarchy.
This mapping is used by the CPU layer of our system to translate brick IDs generated on the GPU to brick requests that are then forwarded to the server.
This allows our system to address more channels and to exchange channels at run-time.

The parameters of our multi-channel page table hierarchy and its brick cache are (1) the brick size in voxels, i.e., the size of a brick in the multi-volume hierarchy and the size of an entry in the brick cache, (2) the brick cache size that determines the maximum working set, and (3) the number $m$ of channels which can be referenced simultaneously.

\subsection{Residency octree nodes}
\label{sec:octree}

\begin{figure}[tb]
 \centering
 \includegraphics[width=\columnwidth]{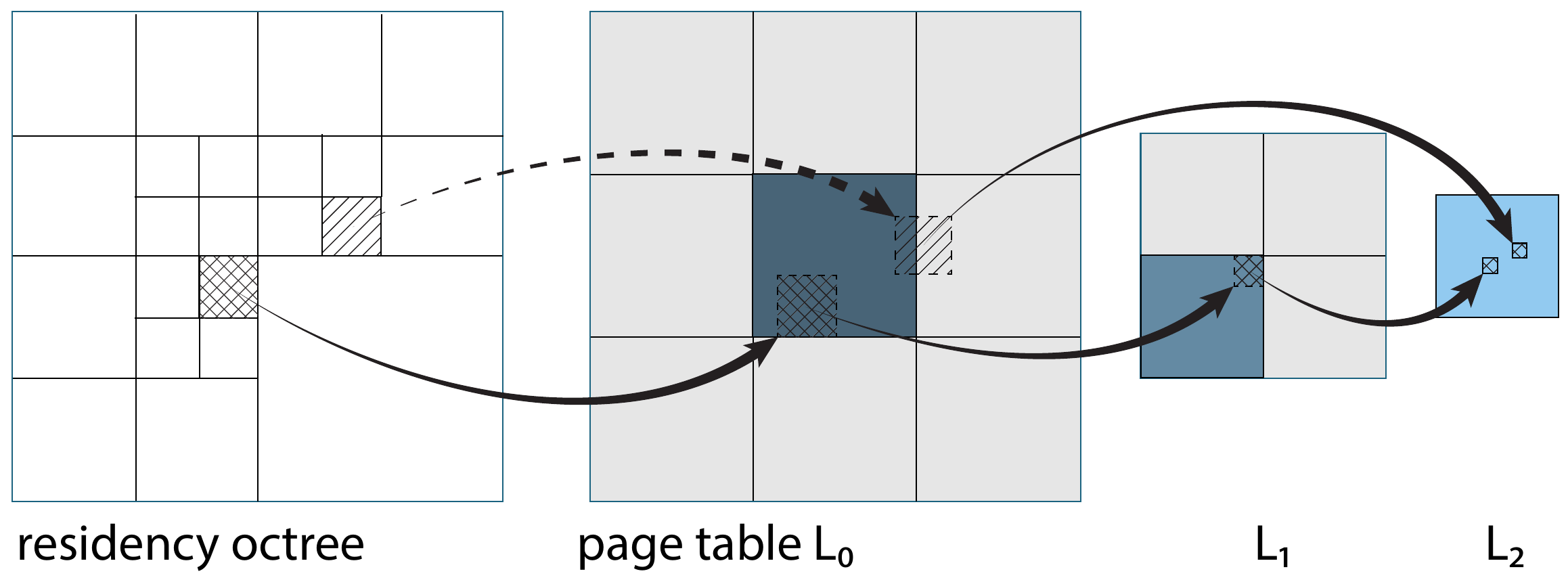}
 \caption{\textbf{Fully vs.\ partially mapped octree nodes.} Blue bricks in the page table (right) are resident in the cache. Dashed lines indicate that a node from the residency octree is \emph{partially mapped}, and solid lines indicate that a node is \emph{fully mapped}. The cross-hatched node in the octree (left) is fully mapped in all resolution levels, $l_0$, $l_1$, and $l_2$, in the page table (right). The single-hatched node is partially mapped in $l_0$, and fully mapped in $l_2$. Note that $l_0$, $l_1$, and $l_2$ cover the same spatial extent.}
\vspace{-4mm}
  \label{fig:fully-partially-mapped}
\end{figure}

For each subdivision level, a residency octree node can be (1) fully mapped, (2) partially mapped, or (3) not mapped at all, as illustrated in \cref{fig:fully-partially-mapped}.
However, since we want to avoid having to load all bricks of a given resolution level when we only need a subset, we only keep track of each node's partial mapping.
Furthermore, each node represents a spatial region in the volume whose corresponding data contain a range of scalar values.
Transfer-function independent metadata for this range, e.g., the minimum and maximum or a histogram of occurrences, is used to determine if a node is empty or homogeneous in the current view.
In these cases, a node's corresponding bricks do not need to be resident in the cache.
If a node is empty or homogeneous, it is implicitly considered fully mapped in all resolution levels during octree traversal.

Each node in the residency octree stores (1) transfer-function independent metadata that can be used for empty-space skipping, (2) pointers to its children, and (3) a bitmask storing in which resolution levels the node is at least partially mapped.
Since the residency octree only stores metadata and information about cache residency, it is not necessary to load successive resolution levels or to load them as a whole.
Instead, individual bricks can be loaded as needed while the tree is constructed and updated incrementally for those regions that are of interest as determined by our ray-guided renderer.
Since the metadata stored in octree nodes is not tied to a specific brick, only the residency information of a node needs to be updated when a brick is added to or removed from the brick cache (\cref{sec:octree-updates}).

To extend the residency octree to multi-channel data, all channels share the same octree structure, but each node stores culling and residency information for up to $m \le n$ channels, where $m$ is the number of channels that can be referenced simultaneously in each octree node.
Using the same octree structure for all channels while keeping a node's residency information independent for all channels makes it possible to use the same spatial subdivision for empty-space skipping, while at the same time rendering each channel in a different resolution.

Since the set of visible channels may change at run-time and volume data may be streamed in on a per-channel basis rather than loading all channels at once, a node may already store culling and residency information for one channel while this information is still missing for other channels.
Therefore, each node also stores for each of the $m$ channels whether the channel information it contains is already initialized or not using a special \texttt{INVALID} value.
If a node's data for a channel is invalid, it stores a pointer to the node in the next lower subdivision level that has valid information for that channel, or a special \texttt{UNKNOWN} value if no such node exists.
To reduce the memory required for each node, the pointer to the next node storing valid information replaces the metadata stored for that channel.
Whenever a channel is replaced by another one, all nodes storing valid data for the old channel are marked as \texttt{INVALID} for the new channel.

The parameters of our multi-channel residency octree are (1) the maximum depth of the tree, and (2) the maximum number of channels that can be referenced by the octree and its underlying multi-channel page table hierarchy (\cref{sec:memory-management}).

\subsection{Residency octree updates}
\label{sec:octree-updates}

Based on the current viewing parameters and the residency octree nodes, a ray-guided renderer determines the metadata and bricks that need to be requested from the server~(\cref{sec:mixed-resolution-multi-volume-renderer}), which implicitly determines when the tree needs to be refined. Whenever new data is streamed in from the server, the residency octree needs to be updated.

\textbf{Updating culling information.}
As soon as new metadata for a volume region has been received by the client, the corresponding residency octree node needs to be either (1) created and added to the octree, or (2) updated if it already exists.
Whenever a new node is added to the residency octree, its metadata is initialized as \texttt{INVALID} for all channels except for the channel for which the metadata was received.
The new node's residency information is also only initialized for this channel, by checking for each resolution level whether the new node is partially mapped or not.
If the node already existed in the residency octree, its channel-specific metadata is updated and its residency information is initialized similarly.
This only happens for multi-channel volumes (\cref{sec:octree}) when a node already holds metadata for another channel.

In case the server does not support requesting metadata for a region in the volume, e.g., if it is just a file server, this information has to be computed on the fly from the actual brick (voxel) data on the client.
Since a residency octree node is not tied to a specific resolution, it is up to the application to choose a resolution level for which the corresponding set of bricks should be fetched from the server in order to have sufficient data to compute the metadata.
To avoid cases where too few voxels result in inaccurate metadata, a residency octree may specify a minimum number of voxels that have to be taken into account when computing the metadata for a node's spatial extent.
If the client had to fetch new brick data in order to compute the node's culling data, each new brick can then also be uploaded to the GPU. This requires residency information to be updated as well.

\textbf{Updating residency information.}
For each new brick received by the client, an available slot in the brick cache is selected for storing it.
Whenever a new brick is stored, we determine all residency octree leaf nodes whose spatial extent overlaps the spatial extent of the brick to mark them as partially mapped in the brick's resolution level and channel.
This update is then propagated up the tree to the root node.

Residency information is stored as bitmasks, and can thus be updated for all non-leaf nodes via bit-wise \texttt{OR} of the bitmasks of child nodes.
This is possible because residency octree nodes store partial residency information and therefore only require a single brick of a resolution level to be resident in the brick cache in order to be partially mapped in that resolution level.
As soon as a node's bitmask remains unchanged by this operation, we terminate the upwards propagation of the update.

\textbf{LRU cache eviction.}
If no cache slot is available for a new brick, we use an LRU scheme~\cite{hadwiger_interactive_2012,sarton_interactive_2020} to evict the least recently accessed brick to create a slot.
After a brick is evicted from the cache, we check for all leaf nodes whose spatial extent overlaps the brick's spatial extent if they are still partially mapped after the evicted brick is removed.
This is done by checking for each of a node's corresponding bricks in the resolution level that the evicted brick belongs to, if they are resident in the cache or not.
Only if no other brick in this resolution level is resident in the cache, the node is no longer partially mapped in that resolution level and the corresponding bit in the node's bitmask is reset.
Non-leaf nodes are updated in the same way as newly added bricks, by setting their bitmasks to the bit-wise \texttt{OR} of their child nodes' bitmasks.

\section{Mixed-Resolution Multi-Volume Rendering}
\label{sec:mixed-resolution-multi-volume-renderer}

\begin{figure*}[tb]
 \centering
 \begin{subfigure}[]{0.23\textwidth}
  \centering
  \includegraphics[width=\textwidth]{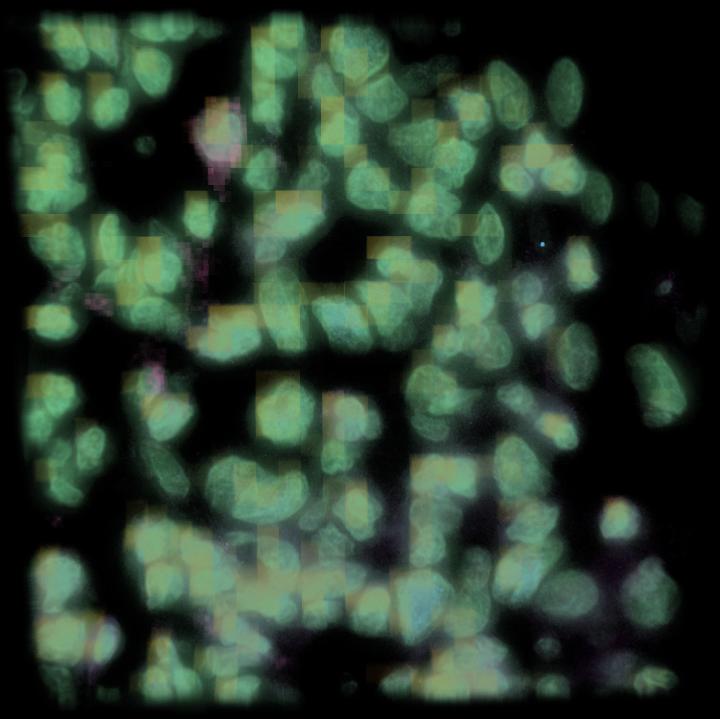}
  \caption{pink $= l_3$, orange $= l_5$, rest $= l_0$}
  \label{fig:mixed-1}
 \end{subfigure}
 \begin{subfigure}[]{0.23\textwidth}
  \centering
  \includegraphics[width=\textwidth]{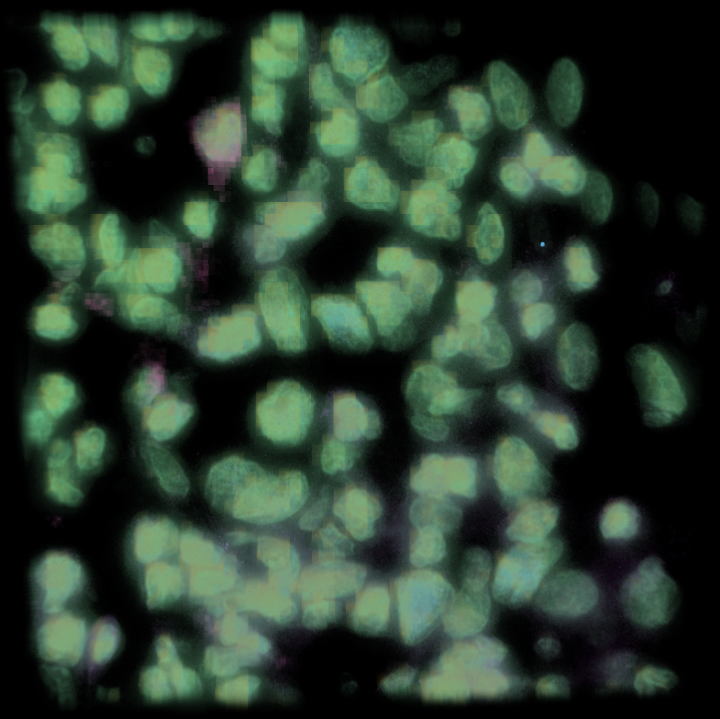}
  \caption{pink $= l_3$, orange $= l_4$, rest $= l_0$}
  \label{fig:mixed-2}
 \end{subfigure}
 \begin{subfigure}[]{0.23\textwidth}
  \centering
  \includegraphics[width=\textwidth]{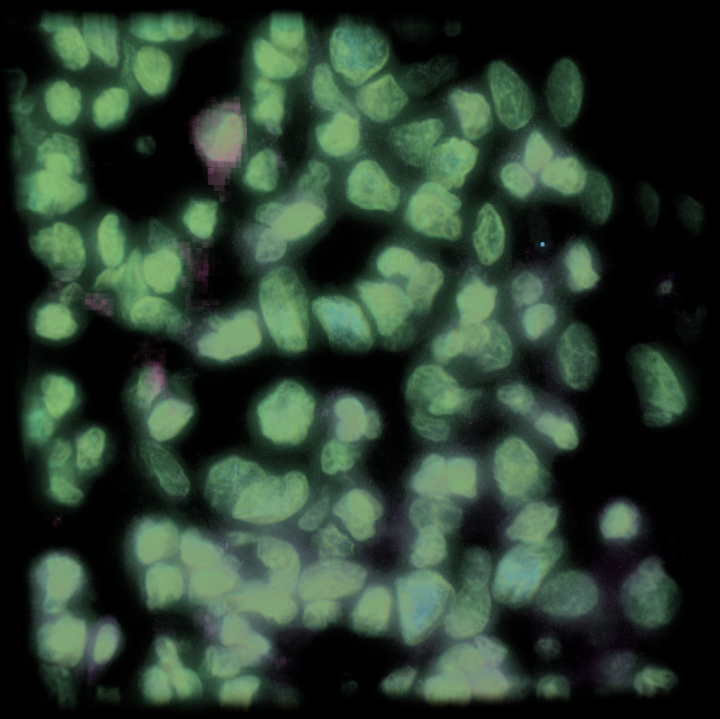}
  \caption{pink $= l_3$, rest $= l_0$}
  \label{fig:mixed-3}
 \end{subfigure}
 \begin{subfigure}[]{0.23\textwidth}
  \centering
  \includegraphics[width=\textwidth]{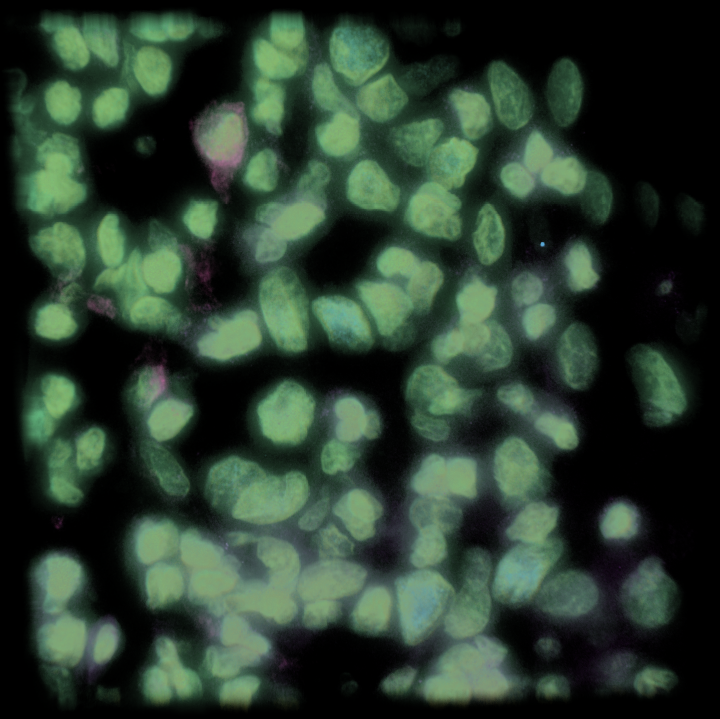}
  \caption{all $= l_0$}
  \label{fig:mixed-4}
 \end{subfigure}
 \vspace{-2mm}
 \caption{\textbf{Mixed-resolution rendering of five channels of human tissue data.} The pink and orange channels are rendered at lower resolutions than the other channels, which are all rendered at the highest resolution level ($l_0$). The pink channel is rendered at resolution level $l_3$ (a-c), and the orange channel is rendered at resolution levels $l_5$ (a), $l_4$ (b), and $l_0$ (c,d). In (d) all channels are rendered at the highest resolution.}
 \label{fig:mixed-resolution-rendering}
\end{figure*}

We use ray-guided volume rendering~\cite{crassin_gigavoxels_2009} to determine the data needed for rendering.
For each pixel, we march along its viewing ray, and at each step traverse the residency octree (\cref{sec:ray-traversal}) to find the leaf node containing the current sample.
From the leaf node metadata, we then determine if the sample falls into an empty region that can be skipped, or if it might contribute to the output image.
In the latter case, the node's residency information is used to access the voxel data of a brick that is resident in the cache.
If the leaf node's metadata is missing or invalid, or the desired brick is not resident in the cache, the brick data is requested from the server.
To save memory when visualizing multiple channels, we support controlling the resolution levels for rendering on a per-channel basis (\cref{sec:mixing-resolutions}), as illustrated in~\cref{fig:mixed-resolution-rendering}.

\subsection{Residency octree traversal}
\label{sec:ray-traversal}

\SetKw{And}{and}
\SetKw{Break}{break}
\SetKw{Continue}{continue}
\SetKwComment{Comment}{/* }{ */}
\SetKwFor{For}{for (}{) $\lbrace$}{$\rbrace$}

\begin{algorithm}[tb]
 \caption{Residency octree traversal. For each sample along the ray, we traverse the tree until we reach a maximum traversal depth or we find an empty node that allows us to skip over empty space. Missing metadata and bricks are reported so they can be streamed in from the server.}
 \label{alg:octree-traversal}
 resolutionLevel = chooseResolutionLevel(depth(ray))\;
 traversalDepth = chooseTraversalDepth(stepSize(ray))\;
 node = rootNode\;
 \For{\upshape $i = 0$; i $\leq$ traversalDepth; i$++$}{
  \If{\upshape isEmpty(node)}{
   ray = skipNode(node, ray)\;
   \Break
  }
  \If{\upshape isHomogeneous(node)}{
   ray = renderAndSkipNode(node, ray)\;
   \Break
  }
  \If{\upshape not(isPartiallyMapped(node))}{
   reportBrickRequest(node, ray, resolutionLevel)\;
   ray = skipNode(node, ray)\;
   \Break
  }
  \If{\upshape i $<$ traversalDepth}{
   nextNode = next(node, ray)\;
    \eIf{\upshape nextNode.missing}{
     reportNodeRequest(nextNode)\;
    }{
     node = nextNode\;
     \Continue
   }
  }
  brick = getBrick(node, ray, resolutionLevel)\;
  \If{\upshape brick.missing}{
   reportBrickRequest(node, ray, resolutionLevel)\;
   brick = getAlternativeBrick(node, ray, resolutionLevel)\;
  }
  \If{\upshape not(brick.missing)}{
   renderUntilNextBoundary(brick, node, ray)\;
  }
  \Break
 }
\end{algorithm}

Traversal of the residency octree to find the node to be sampled for a single channel is outlined in \cref{alg:octree-traversal}.
For brevity, we assume that the root node already contains metadata.
Before traversing the residency octree, we choose the desired resolution level for the ray sample based on current viewing parameters such as the sample's distance to the camera.
Furthermore, we choose a maximum traversal depth based on the current sampling step size, to avoid skipping a distance smaller than the distance to the next ray sample.
The tree is only traversed until this maximum traversal depth is reached.
If a node is empty, all samples along the ray that would fall into the empty node are skipped.
Similarly, if a node is homogeneous, i.e., all voxels hold (approximately) the same value, the node's value is used directly for rendering for all samples along the ray that would fall into the node, and the ray is advanced to the next node.
If a node is not even partially mapped, we can stop traversal and report a brick request.
We request a new node and thus refine the residency octree if the current subtree is not deep enough to reach the target traversal depth and the culling information currently available does not allow skipping the current node.
If the target traversal depth is reached and the current node is non-empty and inhomogeneous, the brick of the chosen resolution level is retrieved from the brick cache.
If this brick is not resident in the cache, a brick request is reported and we try to load a brick from an alternative resolution level in which the node is currently partially mapped.
If no brick is found, we advance the ray to the next sample.
Depending on the resolution level of the brick, the sampling distance along the ray can potentially be increased.

\begin{figure*}[t!]
 \centering
 \includegraphics[width=0.89\textwidth]{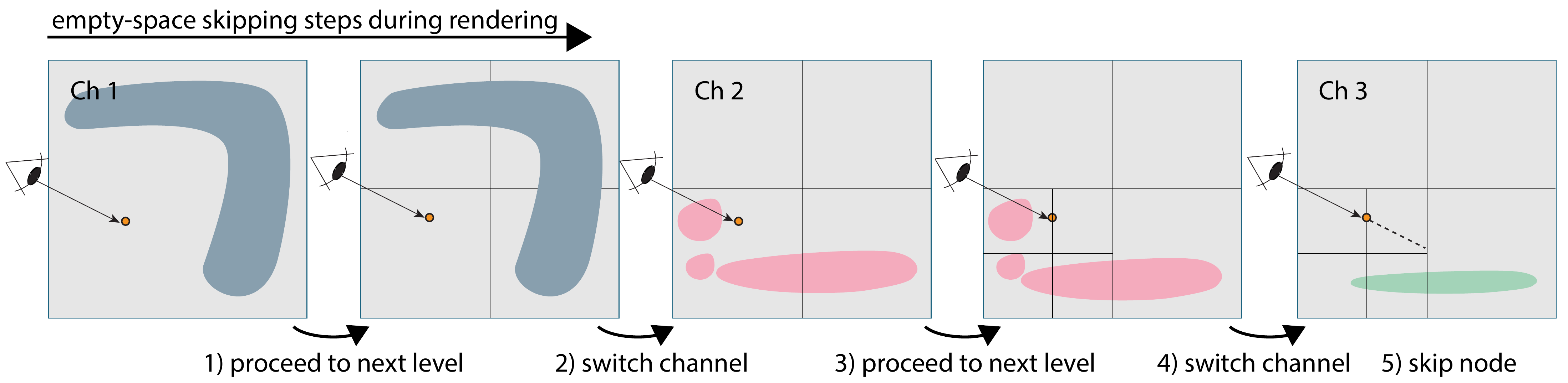}
 \vspace{-1mm}
 \caption{\textbf{Empty-space skipping for multiple channels.} Traversal starts at the root node and the first channel. The node is non-empty, so we proceed to the next subdivision level (1). For the first channel, this node is empty, so we can stop traversing the tree for this level and proceed to the next channel (2). The node is non-empty for channel 2, so we proceed to the next subdivision level (3). We repeat this procedure until we finally can skip the largest node in which all channels are empty (5). }
\vspace{-3mm}
  \label{fig:multichannel-traversal}
\end{figure*}

The algorithm for traversing the residency octree for multi-channel data is similar.
The main difference is that we cannot terminate the traversal before all channels have been processed.
The channels are organized based on their importance, which, together with the current viewing parameters, determines both the sampling rate and the desired traversal depth (\cref{sec:mixing-resolutions}).
Traversing the entire tree for each sample and channel would be computationally inefficient.
Thus, we only traverse the tree hierarchy once, starting with the channel to be rendered in the highest resolution, requiring the highest sampling rate.

To do this, the algorithm keeps track of an index in the sequence of channels.
As soon as we reach a point where we would terminate the traversal in the single channel case, we simply stay in the same node but increase this index, and then continue the traversal from there.
This means that empty nodes can only be skipped if all channels have been found to be empty.
Similarly, homogeneous nodes can also only be rendered more efficiently due to their homogeneity if they are homogeneous for all channels.
So while a node might be empty for one channel, another channel might require a few more traversal steps to reach a subdivision level in which the corresponding node is empty.
The skipped distance is determined by the spatial extent of the last empty node we encounter during traversal.
\cref{fig:multichannel-traversal} illustrates an example of how empty-space skipping may require more traversal steps in a multi-volume setting compared to a single-channel setting because multiple channels have to be checked instead of only one.

Note that the sequence of channels that need to be evaluated for a ray sample depends on the chosen sampling rate for each channel (\cref{sec:mixing-resolutions}).
For example, a channel with high-frequency content may require a higher sampling rate while another channel may only need to be sampled at every second step along the ray.

\subsection{Mixing resolutions}
\label{sec:mixing-resolutions}

When rendering large-scale multi-volume data, it may be desirable to render different channels in different resolutions.
This could be due to a channel having a lower frequency content than the others, or simply one channel not being as important as others for the user, possibly depending on the current viewpoint.
Similarly, a volume might have high-frequency content only in some parts of the volume.
With large-scale data requiring out-of-core methods, we can exploit this by limiting the range of resolutions that are uploaded to the GPU on a per-channel basis to both save memory at run-time and reduce the number of samples that need to be evaluated for all channels.
In our system, each channel has a function that constrains the range of resolution levels to choose from, when computing a resolution level for a channel, e.g., based on current viewing parameters like a ray sample's distance to the camera.
Conceptually, this function can be anything, from user-controlled upper and lower bounds for resolution levels to a function taking into account multiple different parameters like frequency content, viewing parameters, and transfer functions.

\section{Implementation}
\label{sec:implementation}
For our implementation, we use a file server serving bricked multi-resolution multi-volume hierarchies as OME-Zarr files, an implementation of OME-NGFF~\cite{moore_ome-ngff_2021}.
All bricks are LZ4-compressed to reduce the amount of data that has to be streamed in from the server.
Our server does not support querying culling metadata for specific regions in the volume, so they have to be computed on the client.

Our client-side application is implemented in Rust and compiled to WebAssembly. It uses the WebGPU API for issuing GPU commands.
Our implementation uses three separate threads: the main thread controls the UI, a render thread that runs our algorithm and handles all CPU-GPU communication, and a brick-loading thread that in turn uses a pool of worker threads to pre-process data fetched from the file server.
Since WebGPU does not yet allow multiple threads to access the same GPU resources~\cite{maxfield_myles_webgpu_2023}, brick data needs to be transmitted from the brick-loading thread to the render thread.
We decompress bricks and convert all volume data fetched from the server to unsigned 8-bit integer data as a pre-processing step.

Our multi-channel page table hierarchy uses unsigned 32-bit integers for brick IDs.
They consist of 24 bits for the spatial offset (8 bits each for the x, y, and z coordinates) of the brick relative to the origin, and 8 bits for a page table ID uniquely identifying $(m \times k) \leq 256$ page tables, where $m$ is the number of channels and $k$ is the number of resolution levels that can be represented on the GPU.
The brick cache is implemented as a single \texttt{r8unorm} texture.
The brick and cache sizes, as well as $m$ and $k$, are chosen by the user at initialization time.

Our residency octree is implemented as a full, pointerless octree for simplicity.
Each node consists of $m$ unsigned 32-bit integers, where $m$ is the number of channels.
Each channel uses 16 bits for keeping track of partially mapped resolutions, and 8 bits each for minimum and maximum scalar values in the region represented by the node.
The number of subdivision levels is chosen by the user at initialization time.
The multi-channel residency octree is fully implemented on the GPU.

Our mixed-resolution multi-volume renderer is implemented in a WebGPU compute shader.
During ray traversal, we always evaluate each visible channel at each sample to reduce the number of diverging branches.
As an optimization, the renderer never starts tree traversal at the root level, but at a user-defined subdivision level, or the parent level of the subdivision level at which the tree traversal was terminated for the previous sample.
Since we use a full octree, this optimization does not require any changes to our shader code.
Our system's UI has a slider for each channel to control the lower and upper bound for the range of resolutions to choose from during rendering as discussed in \cref{sec:mixing-resolutions}.

\begin{table*}[tb!]
\caption{\textbf{Performance evaluation} of our method for several data sets. We list the general information about the data set and the results of our benchmarks. Note that the data size is the size of the uncompressed volume, not the OME-Zarr data set we converted them to. We compare our method against a multi-resolution multi-channel page table (MRMCPT) only, and an octree-based approach. Bold font is best.}
\vspace{-2mm}
 \centering
\begin{tblr}{
colspec = {|X[c,h]|X[c]|X[c]|X[c]|X[c]|X[c]|X[c]|},
column{1} = {1cm,c},
column{2} = {3.2cm,c},
rows = {font=\small},
row{1}={valign=m}
}
 \hline\hline
 \textbf{Data Set} & \textbf{Description} & \textbf{Data Size and Type, \#~Resolution~Levels} & \makecell{\textbf{DVR Performance}\\Avg. ms / frame (\#~channels)} & \makecell{\textbf{Cache Usage}\\Avg. GPU mem. / frame (\#~channels)}\\ 
 \hline\hline
 \includegraphics[width=0.05\textwidth]{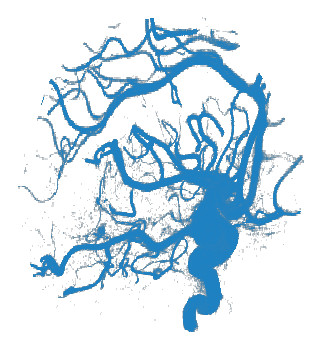}  & \makecell{Aneurism\\$256\times256\times256$\\Channels: 1 (simulated 4)} & \makecell{16.8 MB (1) / 67.2 MB (4)\\8 bit\\Resolution levels: 4} & \makecell{MRMCPT: 5.1 (1) / 15.0 (4)\\Octree: \textbf{4.8} (1) / 15.8 (4)\\Ours: \textbf{4.8} (1) / \textbf{12.3} (4)} & \makecell{MRMCPT: 9.3 MB (1) / 37.3 MB (4)\\Octree: 7.4 MB (1) / 26.4 MB (4)\\Ours: \textbf{5.7 MB} (1) / \textbf{19.9 MB} (4)}\\ \hline
 \includegraphics[width=0.05\textwidth]{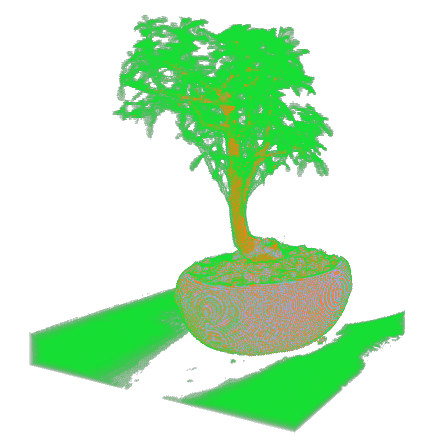} & \makecell{Bonsai\\$256\times256\times256$\\Channels: 1 (simulated 4)} & \makecell{16.8 MB (1) / 67.2 MB (4)\\8 bit\\Resolution levels: 4} & \makecell{MRMCPT: 4.7 (1) / 11.9 (4)\\Octree: \textbf{4.5} (1) /6.7 (4) \\Ours: 4.6 (1) / \textbf{6.5} (4)} & \makecell{MRMCPT: 14.3 MB (1) / 54.8 MB (4)\\Octree: 9.1 MB (1) / 21.0 MB (4)\\Ours: \textbf{7.5 MB} (1) / \textbf{16.5 MB} (4)}\\
 \hline
 \includegraphics[width=0.05\textwidth]{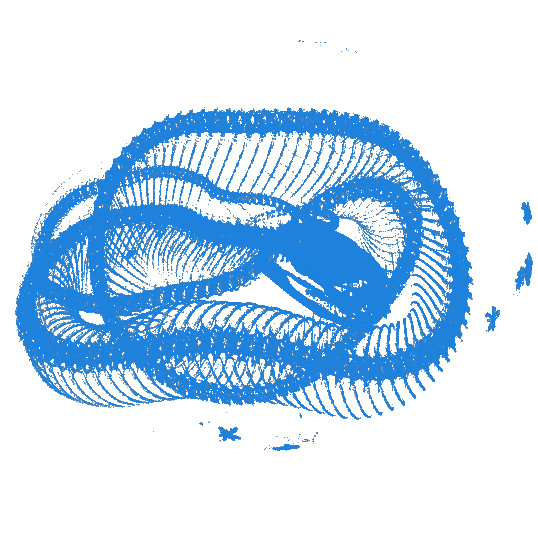} & \makecell{Kingsnake\\$1024\times1024\times795$\\Channels: 1 (simulated 4)} & \makecell{833.6 MB (1) / 3.33 GB (4)\\8 bit\\Resolution levels: 6}  & \makecell{MRMCPT: 15.0 (1) / 42.5 (4)\\Octree: 11.6 (1) / 60.7 (4)\\Ours: \textbf{5.1} (1) / \textbf{19.8} (4)} & \makecell{MRMCPT: 0.66 GB (1) / 1.88 GB (4)\\Octree: 0.23 GB (1) / 0.89 GB (4)\\Ours: \textbf{0.13 GB} (1) / \textbf{0.48 GB} (4)}\\
 \hline
 \includegraphics[width=0.05\textwidth]{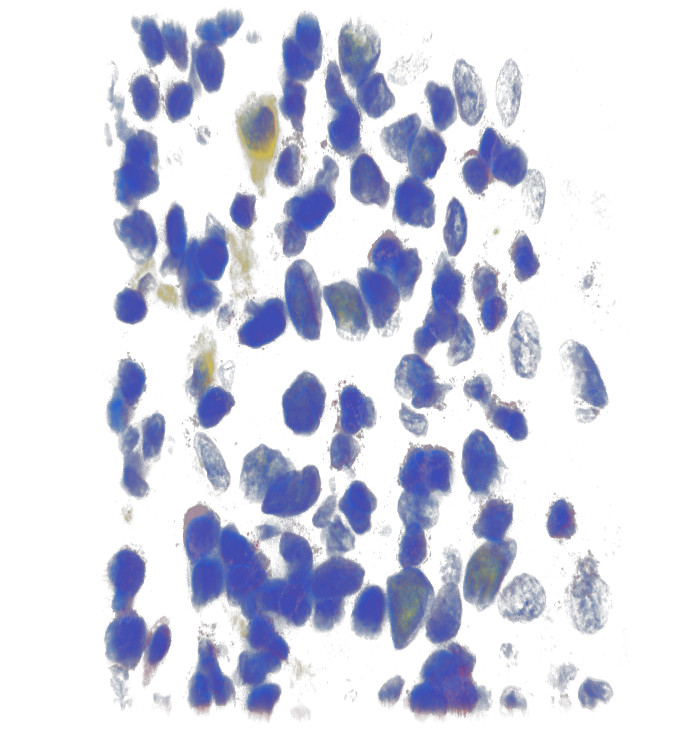} & \makecell{CyCIF Small\\$1024\times1024\times40$\\Channels: 29} & \makecell{2.4 GB\\16 bit\\Resolution levels: 5}& \makecell{MRMCPT: 34.0 (16)\\Octree: 122.8 (16)\\Ours: \textbf{25.7} (16)} & \makecell{MRMCPT: 0.95 GB (16)\\Octree: 0.46 GB (16)\\Ours: \textbf{0.29 GB} (16)}\\
 \hline
 \includegraphics[width=0.05\textwidth]{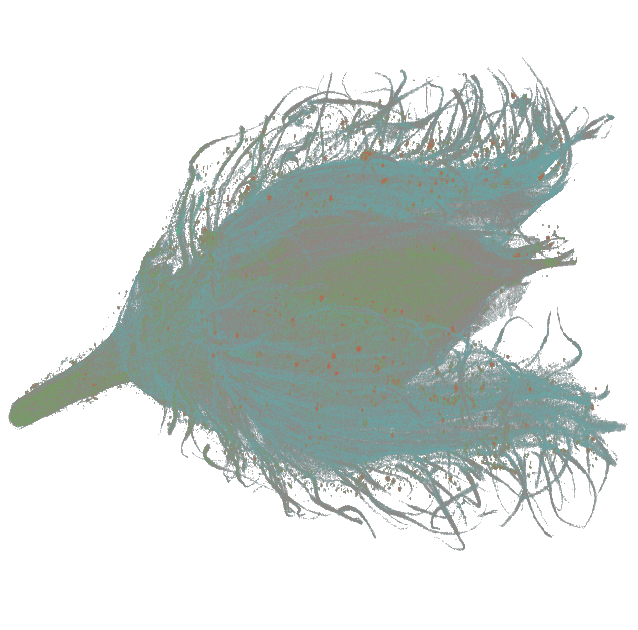} & \makecell{Beechnut\\$1024\times1024\times1546$\\Channels: 1 (simulated 4)} & \makecell{3 GB (1) / 12 GB (4)\\16 bit\\Resolution levels: 6}  & \makecell{MRMCPT: 22.4 (1) / 63.6 (4)\\Octree: 24.8 (1) / 84.4 (4)\\Ours: \textbf{9.5} (1) / \textbf{23.6} (4)} & \makecell{MRMCPT: 0.20 GB (1) / 0.71 GB (4)\\Octree: 0.08 GB (1) / 0.23 GB (4)\\Ours: \textbf{0.06 GB} (1) / \textbf{0.17 GB} (4)}\\
 \hline
 \includegraphics[width=0.05\textwidth]{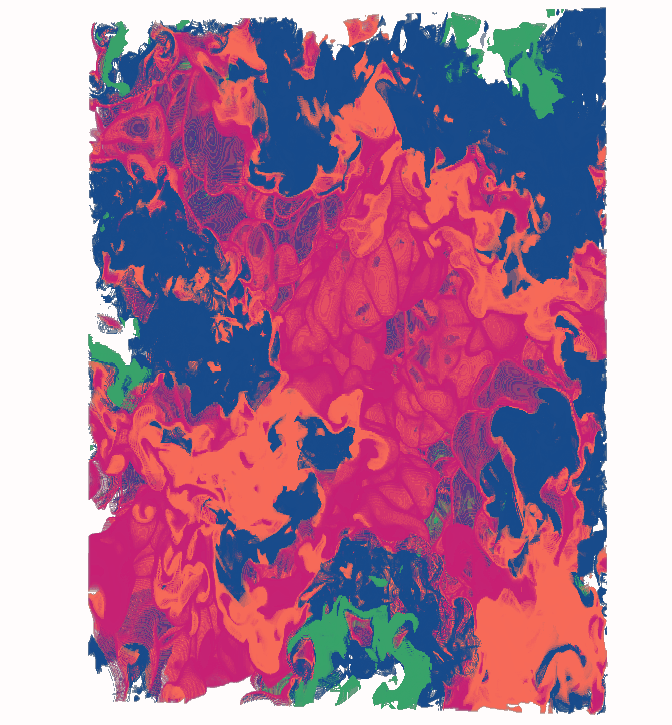} & \makecell{Rayleigh-Taylor Instability~\cite{cook_mixing_2004}\\$1024\times1024\times1024$\\Channels: 1 (simulated 4)} & \makecell{4 GB (1) / 16 GB (4)\\32 bit\\Resolution levels: 5}  & \makecell{MRMCPT: 11.2 (1) / 24.3 (4)\\Octree: 13.8 (1) / 54.9 (4)\\Ours: \textbf{6.4} (1) / \textbf{20.0} (4)} & \makecell{MRMCPT: 0.13 GB (1) / 0.31 GB (4)\\Octree: 0.06 GB (1) / 0.19 GB (4)\\Ours: \textbf{0.04 GB} (1) / \textbf{0.14 GB} (4)}\\
 \hline
 \includegraphics[width=0.05\textwidth]{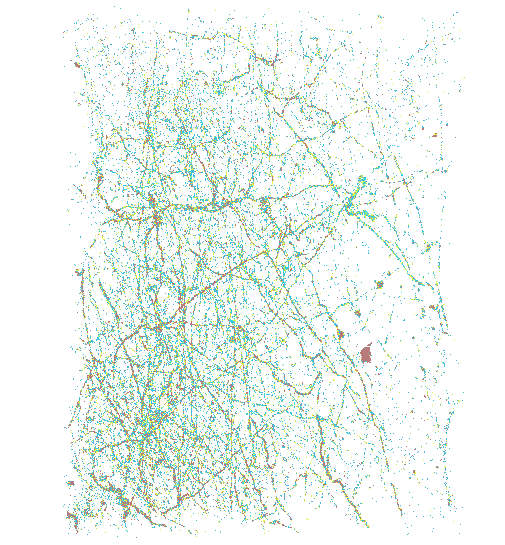} & \makecell{3DNeurons15Sept2016\\$2048\times2048\times1718$\\Channels: 1 (simulated 4)} & \makecell{13.4 GB (1) / 53.6 GB (4)\\16 bit\\Resolution levels: 6}  & \makecell{MRMCPT: 14.4 (1) / 58.2 (4)\\Octree: 52.0 (1) / 221.4 (4)\\Ours: \textbf{14.0} (1) / \textbf{37.9} (4)} & \makecell{MRMCPT: 0.12 GB (1) / 0.47 GB (4)\\Octree: 0.13 GB (1) / 0.50 GB (4)\\Ours: \textbf{0.11 GB} (1) / \textbf{0.43 GB} (4)}\\
 \hline
 \includegraphics[width=0.05\textwidth]{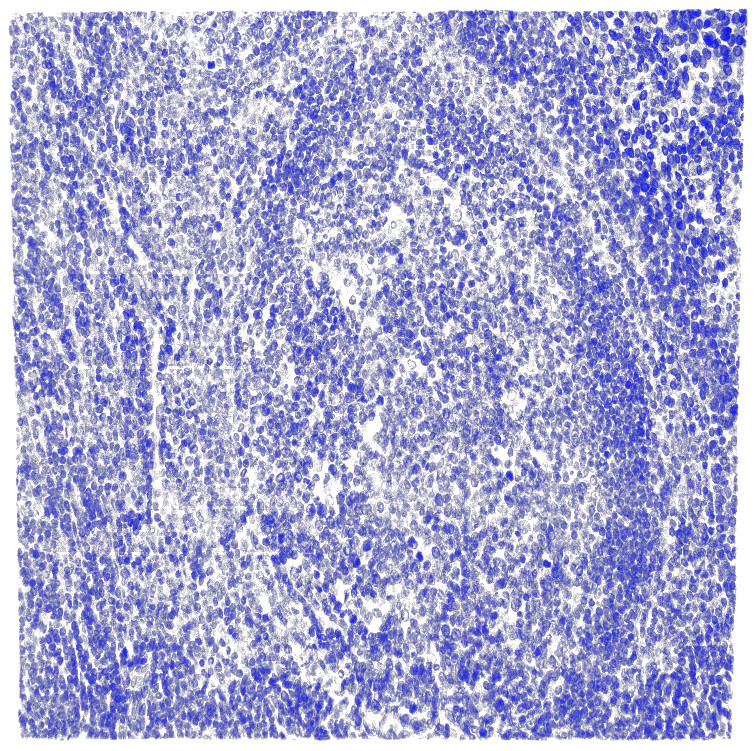} & \makecell{CyCIF Large\\$5632\times4352\times160$\\Channels: 38} & \makecell{149 GB\\8 bit\\Resolution levels: 9} & \makecell{MRMCPT: 24.4 (4)\\Octree: 123.5 (4)\\Ours:  \textbf{22.4} (4)} & \makecell{MRMCPT: 0.24 GB (4)\\Octree: 0.20 GB (4)\\Ours: \textbf{0.17 GB} (4)}\\
 \hline\hline
\end{tblr}
\vspace{-4mm}
\label{tab:results}
\end{table*}

\section{Evaluation and Results}
\label{sec:evaluation}
We evaluate our system on a GeForce GTX 1080 GPU with 8 GB RAM,  AMD Ryzen 7 2700X CPU with 32 GB RAM, using Ubuntu 22.04.1 LTS.
In our experiments, we use Chromium 108.0.5359.40 with the Vulkan backend and experimental WebGPU support enabled.
We evaluate the residency octree and mixed-resolution multi-volume renderer by comparing it to two other approaches: (1) rendering with just a multi-resolution multi-channel page table as described in \cref{sec:memory-management}, and (2) an octree-based out-of-core renderer based on Crassin et al.'s approach~\cite{crassin_gigavoxels_2009} extended to multiple channels implemented by us.

The page table approach (1) directly accesses volume data through the page table and skips empty space only if an entire brick is empty in all channels.
Since bricks are often larger than the spatial extent of octree nodes, empty-space skipping is quite inefficient.
Furthermore, this approach does not substitute missing bricks with volume data from other resolutions that are resident in the cache but skips all samples that fall into the missing brick.
However, for dense volumes, this approach is expected to achieve higher or at least comparable frame rates when compared to the other two approaches since the overhead of traversing a tree cannot be compensated by skipping over empty space.

The octree approach (2) uses the same culling data as our residency octree but uses a one-to-one mapping of bricks and octree nodes.
In order to render parts of the volume at high resolution, whole subtrees have to be resident in cache, leading to slightly higher memory usage.
This applies to all channels. We note that anisotropic volumes and downsampling ratios increase this problem since the overhead of storing full subtrees is often 1/3 (downsampling in two dimensions) instead of 1/7.
This approach is expected to achieve frame rates comparable to our method for a single channel, since both use a similar empty-space skipping strategy.
However, for multiple channels, the octree has to be traversed for each channel to ensure that all subtrees are cache resident.

For evaluation, we use both open single-channel data sets and biomedical data sets generated by CyCIF.
For single-channel data we also simulate multi-channel data with $n$ channels by duplicating the volume $n$ times, rendering channels with different transfer functions.

\textbf{Rendering performance.}
To evaluate the rendering performance of our mixed-resolution multi-volume renderer, we measure the computation time of the compute pass performing the volume rendering using WebGPU's time-stamp query feature~\cite{maxfield_myles_webgpu_2023}.
We measure the performance in 10-second intervals during which the camera is rotating around the center about the y-axis (pointing up), and average the results of 10 iterations each.
To compare the performance of our renderer to the other two approaches, we use the same downsampling ratio and spatial subdivision for both the bricked multi-volume hierarchy and the residency octree.
We use a 4 GB cache and $32^3$ bricks.
All open data sets are tested with one and four channels each, \textit{CyCIF Small}~\cite{nirmal_spatial_2022} is tested with 16 channels, and \textit{CyCIF Large} is tested with four visible channels.
As an optimization, we start residency octree and octree traversal at the third level and do not start traversal at the root for each sample, but from the parent node of the node visited for the last sample.
As to not give an unfair advantage to any of the methods, in most of our benchmarks most bricks visible are already resident in the cache.

Our method outperforms the other approaches in all benchmarks as shown in \Cref{tab:results}.
However, the most significant performance gains are achieved for sparse volumes such as the \textit{Kingsnake}, \textit{Beechnut}, and \textit{3DNeurons15Sept2016} data sets, and for a large number of channels such as for the \textit{CyCIF Small} data set.
For the thin \textit{CyCIF Large} and the dense \textit{Rayleigh-Taylor Instability}~\cite{cook_mixing_2004} data sets, our empty-space skipping method has less of an advantage over the page table approach since the overhead of traversing an octree is not compensated as much by the little empty space that is skipped.
Note that for all data sets, the performance depends on the transfer functions defining if a region can be considered empty.
As expected, the octree-based approach suffers from the computational cost of restarting tree traversal for each channel for multi-channel data sets and increased memory footprint.

\begin{figure*}[tb]
 \centering
 \begin{subfigure}[]{0.30\textwidth}
  \centering
  \includegraphics[width=\textwidth]{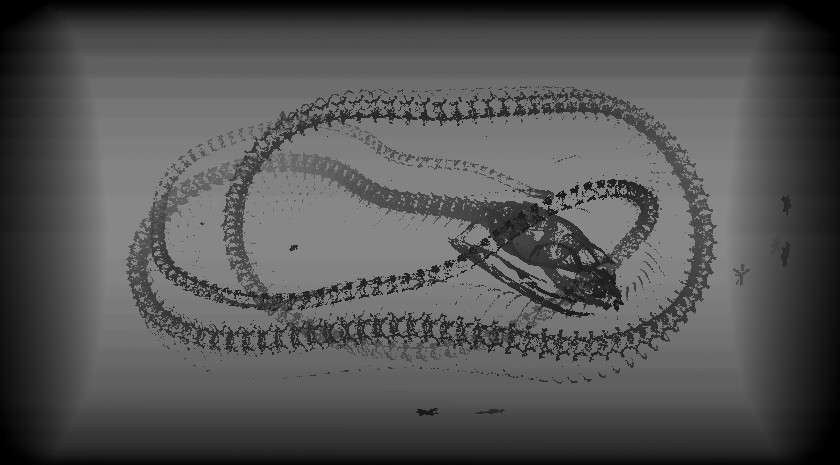}
  \caption{Multi-channel page table hierarchy}
  \label{fig:req-bricks-1}
 \end{subfigure}
 \begin{subfigure}[]{0.30\textwidth}
  \centering
  \includegraphics[width=\textwidth]{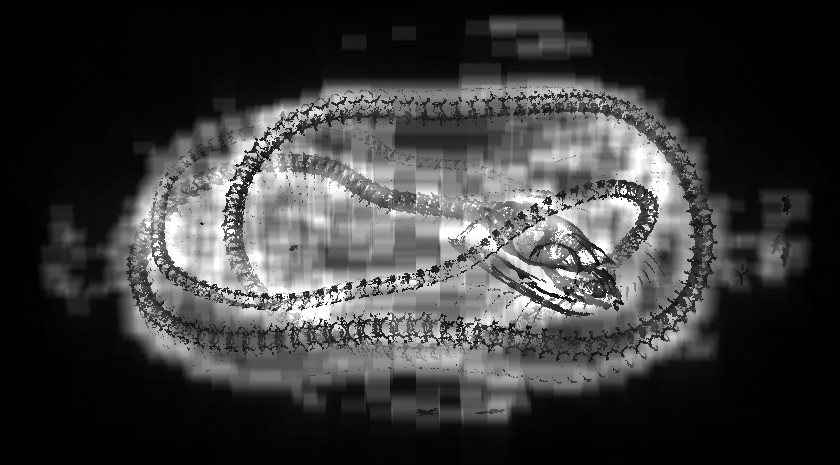}
  \caption{Octree}
  \label{fig:req-bricks-2}
 \end{subfigure}
 \begin{subfigure}[]{0.33\textwidth}
  \centering
  \includegraphics[width=\textwidth]{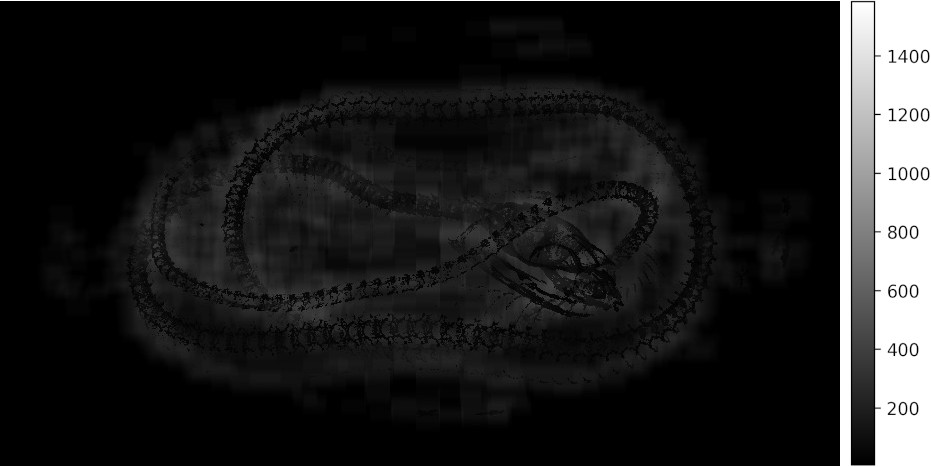}
  \caption{Residency octree}
  \label{fig:req-bricks-3}
 \end{subfigure}
\vspace{-3.5mm}
  \caption{\textbf{Number of bricks required to be resident in the cache per pixel.} Brighter pixels indicate more bricks. (a) Page table hierarchies have limited empty-space skipping capabilities and can access bricks that potentially contain many values outside the currently visible value range. (b) Octrees require lower resolutions in the cache in order to render higher resolutions. (c) Residency octrees do not have either of these limitations, only requiring bricks that are visible under the current viewing conditions to be in the cache, allowing for larger working sets and/or less cache thrashing.}
 \vspace{-5mm}
 \label{fig:required-bricks}
\end{figure*}

\textbf{GPU memory usage.}
We also evaluate the number of bricks each method strictly requires to be resident in the cache per frame under the same viewing conditions.
We report the average amount of GPU memory per frame required by each method over ten 10-second intervals for all data sets in \Cref{tab:results}, showing a consistent improvement of our method in comparison to the two baseline methods.
Page tables only skip entirely empty bricks.
The residency octree supports finer-grained empty-space skipping.
Octrees share this characteristic but require more bricks to be resident in the cache.
Residency octrees reduce the number of required bricks in the cache,
as illustrated in \cref{fig:required-bricks}.

One advantage of not requiring all lower resolutions to be cache resident is a larger high-resolution working set size. This reduces cache thrashing, or vice versa avoids forcing a lower resolution in order to avoid cache thrashing. We note that this can come at the price of not having lower resolutions available for quickly zooming out. However, residency octrees do support both choices, enabling users to choose a suitable trade-off, e.g., either higher-resolution rendering of the current view or fast zooming without needing to page in lower resolutions.

\begin{figure}[b]
\vspace{-0.3mm}
 \centering
 \includegraphics[width=\columnwidth]{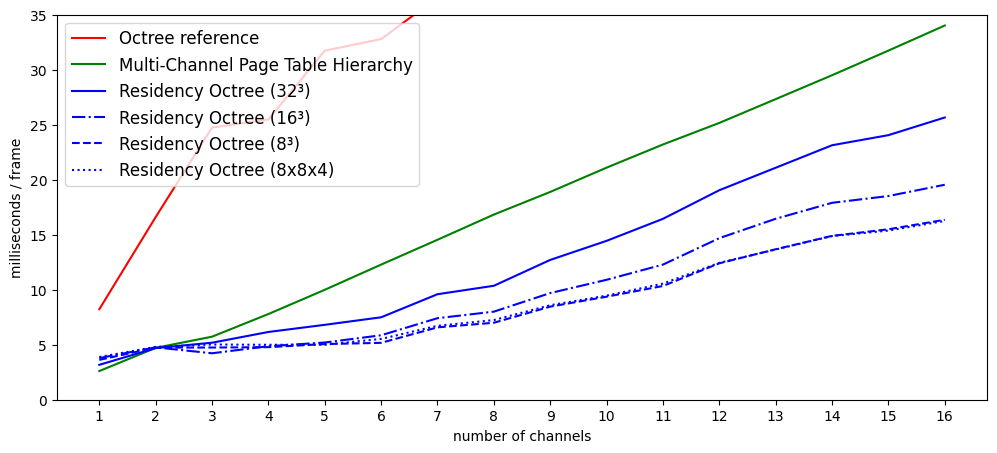}
\vspace{-6.5mm}
 \caption{\textbf{Performance results} for decoupled brick and leaf node sizes for the \textit{CyCIF Small} data set.
 Small residency octree nodes mixed with larger brick sizes achieve the best performance.}
  \label{fig:benchmark}
\end{figure}

\textbf{Decoupling resolution levels from spatial subdivision.}
Residency octrees support finer granularities for empty-space skipping than the bricking granularity used for the voxel data.
\cref{fig:benchmark} shows results for the \textit{CyCIF Small} data set using different spatial subdivisions and different numbers of visible channels.
This volume is very thin and does not contain many empty  $32^3$ bricks.
In this case, our method performs similarly to accessing cached volume data through the multi-channel page table hierarchy directly when visualizing up to three channels.
However, for larger numbers of channels, our method clearly outperforms the other two methods.
When using a more fine-grained spatial subdivision for culling than the bricking granularity, the advantages of the residency octree become even more apparent.
The residency octree also supports completely different spatial subdivision schemes that are better suited for anisotropic volumes, e.g., where a leaf node roughly corresponds to $8 \times 8 \times 4$ voxels, as also demonstrated in \cref{fig:benchmark}.

\section{Discussion and Limitations}
\label{sec:discussion}

\subsection{Limitations}
The memory required for storing the residency octree scales linearly with the number of channels that can be visualized at the same time, because residency information is stored in the same way in each node for each channel (one bitmask per channel).
This is problematic for data sets with a large spatial extent and many channels to be visualized at the same time.
Combining multiple channels into a single channel (e.g., by using dimensionality reduction techniques) could help avoid this issue.
Additionally, for cache coherency, storing multiple channels in an interleaved manner, e.g., using a four-component texture format, as well as using compressed texture formats, might be beneficial.

Furthermore, in our implementation, we currently let the user set the importance of each channel.
We leave investigating different methods to automatically determine the resolution range per channel and region in the volume based on its frequency content to future work.

\subsection{Discussion}

\textbf{Residency octrees combine advantages of page tables and octrees.} 
Our method shares characteristics of existing out-of-core volume rendering methods such as page tables and octrees, but combines their advantages while avoiding their disadvantages.
Like page-table-based approaches, residency octrees support direct access to volume bricks of a desired resolution instead of having to keep lower resolutions resident in the cache unnecessarily.
In case of cache misses, residency information stored in residency octree nodes is used to substitute missing bricks with bricks from another resolution level.
This avoids unnecessary texture lookups for resolutions that are guaranteed not to be resident in the cache.
Similar to octree-based approaches, our tree data structure supports more efficient empty-space skipping than page tables.
But by decoupling the resolutions in the data set from the spatial subdivisions determined by the tree, our approach allows for more fine-grained empty-space skipping than previous approaches do while at the same time being more flexible in terms of data access patterns.

\textbf{Flexible mixing of different resolutions.} In comparison to other methods, residency octrees are more flexible when mixing different resolutions while also skipping empty space in an efficient manner.
Our method not only makes it possible to mix resolutions when rendering a single channel but also when rendering multiple channels.
This makes it feasible for streaming in multi-channel data from a remote server on demand and efficiently switching out channels at run-time.

\textbf{Efficient multi-channel rendering.} 
By traversing the octree only once for each sample, we minimize the performance cost of traversing a hierarchy when rendering multi-volume data.
Residency octrees are constructed incrementally, i.e., subtrees that are never visible on screen are never constructed.
Residency octrees and our mixed-resolution multi-volume algorithm are more efficient than previous approaches for a large number of visible channels, especially when optimizing the tree's subdivision for empty-space skipping.
Furthermore, residency octrees require fewer bricks to be resident in the cache than previous approaches and thus optimize cache usage for large-scale rendering.

\section{Conclusions and Future Work}
\label{sec:conclusion}
In this work, we have presented the \emph{Residency Octree}, a hybrid data structure (multi-resolution page tables combined with a special octree data structure) that is well-suited for client-side web-based out-of-core volume rendering of data sets with a large number of channels.
This data structure decouples the cache residency of multi-resolution data from a resolution-independent spatial subdivision determined by the tree.
This makes it possible to efficiently and adaptively choose and mix resolutions, adapt sampling rates, and compensate for cache misses by rendering other resolutions that are resident in the cache.
At the same time, this decoupling allows residency octrees to support fine-grained empty-space skipping, independent of the data subdivision used for caching. This is beneficial in web-based scenarios where the bricking granularity may be under the control of a third party.
In the future, we intend to experiment with other spatial subdivision schemes to exploit the benefits of decoupling data resolutions and subdivisions further.

\section*{Acknowledgments}
This work was funded by the NCI award U2C-CA233262 and the NSF award IIS-1901030. It has been supported in part by King Abdullah University of Science and Technology (KAUST). The paper was partly written in collaboration with the VRVis
Competence Center. VRVis is funded by BMVIT, BMWFW, Styria, SFG, and
Vienna Business Agency in the scope of COMET–Competence Centers for
Excellent Technologies under Grant 854174 which is managed by FFG. We especially thank Clarence Yapp, who recorded and registered the 3D immunofluorescence data we use for the performance evaluation.

\bibliographystyle{abbrv-doi-hyperref}

\bibliography{ms}

\appendix 

\end{document}